\documentclass[doublespacing]{elsart}

\usepackage{epsf}

\usepackage{amssymb}

\begin{document}

\begin{frontmatter}

% use the thanksref command within \title, \author or \address for 
%footnotes;
% use the corauthref command within \author for corresponding author 
%footnotes;

\title{Nonequilibrium Dynamics of Coupled Quantum Systems}
\author[ad1]{G. Flores-Hidalgo \corauthref{cor1}},
\ead{gflores@cbpf.br}
\author[ad2]{Rudnei O. Ramos} 
\ead{rudnei@uerj.br}

\corauth[cor1]{Corresponding author.}

\address[ad1]{ Centro Brasileiro de Pesquisas Fisicas-CBPF,
Rua Dr. Xavier Sigaud 150, 22290-180 Rio de Janeiro, RJ, Brazil}

\address[ad2]{Departamento de F\'{\i}sica 
Te\'orica,
Universidade do Estado do Rio de Janeiro,
20550-013 Rio de Janeiro, RJ, Brazil }

\begin{abstract}

The nonequilibrium dynamics of coupled quantum oscillators
subject to different time dependent quenches are analyzed
in the context of the Liouville-von Neumann approach.
We consider models of quantum oscillators in interaction
that are exactly soluble in the cases of both sudden and
smooth quenches. The time evolution of number densities and the 
final equilibration distribution for 
the problem of a quantum oscillator coupled to an infinity 
set of other oscillators (a bath) are explicitly worked out.

\end{abstract}

\end{frontmatter}   

% main text starts here like in revtex

\section{Introduction}

The nonequilibrium dynamics of quantum systems are a timely and
relevant subject that is important to many areas of study in physical
science, from condensed matter systems to cosmology. In most cases we
are usually interested in the processes of thermalization and
equilibration, the determination of the relevant time scales involved,
together with an understanding of the generation of entropy and particle
production in nonequilibrium dissipative systems interacting with an
environment. However, despite the importance associate to these
processes, nonequilibrium problems are still poorly understood (see,
{\it e.g.} \cite{reviews}). The nontrivial nonequilibrium dynamics of
fields, for instance, have diverse applications, finding use {\it e.g.}
in the studies concerning the recent experiments in ultra-relativistic
heavy-ion collision \cite{DCC}; applications to the current
problems of parametric resonance and particle production in cosmology
\cite{lindereh}; or in the context of the recent studies involving the
intrinsic dissipative nature of interacting fields \cite{GR,BGR,BR}. In
addition to that, typical problems we have in mind to study are those
related to the nontrivial out-of-thermal equilibrium dynamics
associated with phase transitions in different physical systems. As a
few examples we may cite include the current applications to the study of
formation of Bose-Einstein condensates after a temperature quench
\cite{BEC}, or in the study of the dynamics of coupled fields displaced
from their ground states as determined by their free energy densities
\cite{RF}.

One problem that we will be particularly interested in here is that one
that involves quenched phase transitions, either caused by a sudden
change in temperature of the system, or by a time change of the
parameters of the model (like frequency or couplings), in which case we
have to deal with situations characterized by mixed states involving
general density matrices and where an equilibrium or quasi-equilibrium
formalism do not apply. Common tools used to study such nonequilibrium
quantum systems are the Schwinger's closed time path functional
formalism and the canonical quantization methods \cite{ctpreviews}. In
this paper we will mostly be concerned with the understanding of simple,
exactly soluble quench scenarios from the stand point of the
Liouville-von Neumann (LvN) quantum canonical formalism \cite{Lewis}. In
this formalism we have that the time evolution of the
density operator $\hat{\rho}$ is given by the LvN equation with
Hamiltonian operator $\hat{H}$:

\begin{equation}
\frac{\partial}{\partial t}\hat{\rho} (t)
=i[\hat{\rho}(t), \hat{H}]
\label{lvn1}
\end{equation}
and satisfying the usual normalization condition ${\rm Tr}~\!\hat{\rho} =1$.
In order to solve Eq. (\ref{lvn1}) we can make use of the following
fact.
Suppose any two operators $\hat{o}_1(t)$ and $\hat{o}_2(t)$ that
satisfy the LvN equation ($j=1,2$)

\begin{equation}
\frac{\partial}{\partial t} \hat{o}_j(t)=i[\hat{o}_j(t),\hat{H}],
\label{lvn2}
\end{equation}
then we can easily show that any operator
function of $\hat{o}_1(t)$ and $\hat{o}_2(t)$, 
$f(\hat{o}_1,\hat{o}_2)$, 
will also satisfy Eq. (\ref{lvn2})
provided that the time dependence in $f$ only comes through its
dependence in $\hat{o}_1$ and $\hat{o}_2$. This important property 
allows us
to construct an exact solution to Eq. (\ref{lvn1}) as a function of 
time dependent operators that satisfy the LvN equations. 
The LvN approach has been recently used in Refs.
\cite{Kim1,Kim2} to treat different problems in quantum field theory 
and cosmology and also emphasizing the advantages of the method 
compared 
to other approaches for nonequilibrium dynamics of quantum systems.
The method may also be more efficient than the more common
functional integral method, where most of the time we have to
deal with nonlocal terms in time that can only be appropriated
deal with by recurring to linear response and/or adiabatic approximations
(by requiring that the system moves sufficiently slowly as compared to typical
microscopic time scales) that may not always apply to physically interesting
systems (see for
example Refs. \cite{GR,BGR,BR}).

In this paper we will study the use and applicability of the LvN approach 
for solving different systems of
quantum oscillators subject to different quenches that are exactly
soluble. With the study of these systems we expect to gain enough
insight to then extend this analysis to more complex problems, including
the nontrivial cases involving nonlinear interactions, where usually we
must make use of perturbative methods, like in quantum fields in
interaction and in the description of system evolution (like through a
time dependent order parameter), when interaction with a thermal
environment is included. In the next section we will give a simple
illustration of use of the LvN approach by solving the (frequency) time
dependent harmonic oscillator. Then in Sec. III we will study two
examples of interacting quantum systems composed of two harmonic
oscillators, initially in different thermal states, that are then
coupled linearly at some given time. The case of a smooth quench is
explicitly solved. Then in Sec. IV we study the case of a system, 
represent by an harmonic
oscillator, coupled linearly to a bath, represented by an infinity set of 
other oscillators,
like in Caldeira-Leggett models and we explicitly evaluate  the number
density for the system, in the case of a sudden quench, in the context of 
the LvN
approach. In Sec. V we discuss a general result concerning the addition
of nonlinear interactions, for a model with quartic interactions, in the
context of the approach discussed in the previous sections. {}Finally in
Sec. VI we give our concluding remarks. Two Appendixes are included to
clarify a few technical details.

\section{The Time Dependent Harmonic Oscillator}

Let us begin by showing how the LvN approach works in the simplest
problem of the time dependent harmonic oscillator case, described 
by the time dependent Hamiltonian:

\begin{equation}
\hat{H}=\frac{1}{2} \hat{p}^2 +\frac{1}{2}\omega^2(t)\hat{q}^2 ,
\label{hosci}
\end{equation}
with a time dependent frequency $\omega(t)$. We consider that initially
the system is in thermal equilibrium state at a temperature $T=1/\beta$
(note that for convenience we use throughout this paper the natural units,
$k_B=\hbar=1$).
Next, we introduce time dependent annihilation and creation operators,
$\hat{a}(t)$ and $\hat{a}^\dag (t)$, given by

\begin{equation}
\hat{a}(t)=A(t)\hat{q}+B(t)\hat{p} 
\label{aa}
\end{equation}
and 

\begin{equation}
\hat{a}^\dag(t)=A^\ast(t)\hat{q}+B^\ast(t)\hat{p},
\label{aa1}
\end{equation}
where the time dependence comes through the c-number coefficients
$A(t)$ and $B(t)$. {}From the relation

\begin{equation}
[\hat{a}(t),\hat{a}^\dag(t)]=1 ,
\label{alge}
\end{equation}
we obtain that

\begin{equation}
A(t)B^\ast(t)-B(t)A^\ast(t)=-i.
\label{wrons}
\end{equation}
Using Eq. (\ref{aa}) in Eq. (\ref{lvn2}), we obtain

\begin{equation}
\dot{A}(t)\hat{q}+\dot{B}(t)\hat{p}=-A(t)\hat{p}+\omega^2(t)B(t)\hat{q},
\label{time}
\end{equation}
where dots mean time derivatives. {}From Eq. (\ref{time}) we then get
that

\begin{equation}
\ddot{B}(t)+\omega^2(t)B(t)=0
\label{oscil}
\end{equation}
and  $A(t)$ is related to $B(t)$ by the relation

\begin{equation}
A(t)=-\dot{B}(t).
\label{ssol}
\end{equation}
Using the above equation we can rewrite Eqs. (\ref{aa}),
(\ref{aa1}) and (\ref{wrons}) and express them as

\begin{equation}
\hat{a}(t)=B(t)\hat{p}-\dot{B}(t)\hat{q},
\label{an1}
\end{equation}

\begin{equation}
\hat{a}^\dag(t)=B^\ast(t)\hat{p}-\dot{B}^\ast(t)\hat{q}
\label{an2}
\end{equation}
and

\begin{equation}
\dot{B}(t)B^\ast(t)-B(t)\dot{B}^\ast(t)=i.
\label{wrons1}
\end{equation}

It is easy to see that Eq. (\ref{wrons1}) is the Wronskian of Eq.
(\ref{oscil}), {\it i.e.}, warranting that the operators
$\hat{a}(t)$ and $\hat{a}^\dag(t)$ are annihilation and creations
operators at all times. Now since our annihilation and creation
operators satisfy the LvN equation, any function of them must also
equally satisfy it too. In particular we can construct an exact solution
for $\hat{\rho}(t)$. Since our initial condition is one of thermal
equilibrium we can, therefore, write

\begin{equation}
\hat{\rho}(t)=\frac{1}{Z_\beta}
\exp\left\{-\beta\omega_0\left[\hat{a}^\dag(t)\hat{a}(t)+
\frac{1}{2}\right]\right\},
\label{dens}
\end{equation}
where

\begin{equation}
Z_\beta={\rm Tr}~\exp\left\{-\beta\omega_0
\left[\hat{a}^\dag(t)\hat{a}(t)+\frac{1}{2}\right]\right\}
\end{equation}
and $\omega_0=\omega(t\rightarrow -\infty)$, {\it i.e}, we are
choosing as initial condition that our system is at thermal
equilibrium at $t\rightarrow -\infty$ with temperature $1/\beta$.
Since relation (\ref{alge}) is satisfied, we can define a time
dependent number operator $\hat{N}(t)=\hat{a}^\dag(t)\hat{a}(t)$, 
with time dependent number eigenvectors $ |n(t)\rangle $. 
However,  because Eq. (\ref{alge}) is a time independent relation, 
the number eigenvalues $n$ are time independent, {\it i.e}, we have

\begin{equation}
\hat{N}(t)| n(t)\rangle =n| n(t)\rangle .
\label{eigenv}
\end{equation}

Using Eq. (\ref{eigenv}) it is easy to show that $Z_\beta$ is the
partition function for the initial thermal equilibrium state,

\begin{eqnarray}
Z_\beta&=&\sum_{n=0}^{\infty}\langle n(t)|
 \exp\left\{-\beta\omega_0\left[
\hat{N}(t)+\frac{1}{2}\right]\right\}|n(t) \rangle \nonumber\\
&=&\sum_{n=0}^{\infty} \exp\left\{-\beta\omega_0\left(n+
\frac{1}{2}\right)\right\}\nonumber\\
&=&\frac{1}{2\sinh\left(\frac{\beta\omega_0}{2}\right)}.
\label{parteq}
\end{eqnarray}
Note that $\hat{\rho}(t)$, given by Eq. (\ref{dens}), solves 
exactly the out-of-equilibrium dynamics of the system 
with the condition of thermal equilibrium as
initial state. In order to compute the time dependent thermal
averages of certain operators it is convenient to rewrite Eqs.
(\ref{an1}) and (\ref{an2}) in terms of the initial annihilation and
creation operators, that we will denote simple as $\hat{a}$ and 
$\hat{a}^\dag$. {}For this end we first solve for $\hat{q}$ and 
$\hat{p}$ in terms of the initial values of $B(t)$ and $\dot{B}(t)$ 
that we will denote just as $B$ and $\dot{B}$, respectively, without 
the time argument. We then find that

\begin{equation}
\hat{q}=i(B^\ast\hat{a}-B\hat{a}^\dag),
\label{q1}
\end{equation}

\begin{equation}
\hat{p}=i(\dot{B}^\ast\hat{a}-\dot{B}\hat{a}^\dag).
\label{p1}
\end{equation}
Using now Eqs. (\ref{q1}) and (\ref{p1}) in Eqs. (\ref{an1}) and
(\ref{an2}), we then obtain the Bogoliubov transformation:

\begin{equation}
\left(\begin{array}{c}\hat{a}(t)\\
\hat{a}^\dag(t)\end{array}\right)=
\left( \begin{array}{cc}\alpha(t)&\beta(t)\\
\beta^\ast(t)&\alpha^\ast(t)\end{array}\right)
\left(\begin{array}{c}\hat{a}\\
\hat{a}^\dag\end{array}\right),
\label{bugo}
\end{equation}
where $\alpha(t)$ and $\beta(t)$, the Bogoliubov
coefficients, are given by

\begin{equation}
\alpha(t)=i\left(\dot{B}^\ast B(t)-B^\ast \dot{B}(t) \right),
\label{bug1}
\end{equation}

\begin{equation}
\beta(t)=i\left(B\dot{B}(t)-\dot{B}B(t)\right).
\label{bug2}
\end{equation}
In terms of $\alpha(t)$ and $\beta(t)$
Eq. (\ref{wrons1}) gives then the following relation:

\begin{equation}
|\alpha(t)|^2-|\beta(t)|^2=1.
\label{norm}
\end{equation}

The utility of Eq. (\ref{bugo}) is clear once we compute the thermal expectation
value of the number operator $\hat{N}$,

\begin{eqnarray}
\bar{n}(t)&=&{\rm Tr}
\left[\hat{\rho}(t)\hat{a}^\dag\hat{a}\right]\nonumber\\
&=&\frac{1}{Z_\beta}
\sum_{n=0}^\infty\langle n(t)|
\exp\left\{-\beta\omega_0\left[\hat{a}^\dag(t)\hat{a}(t)+
\frac{1}{2}\right]\right\}\hat{a}^\dag\hat{a}~\! |n(t)\rangle .
\label{numbert}
\end{eqnarray}
Solving Eq. (\ref{bugo}) for $\hat{a}$ and $\hat{a}^\dag$ in terms 
of $\hat{a}(t)$ and $\hat{a}^\dag(t)$, substituting them in 
Eq. (\ref{numbert}) and using Eq. (\ref{eigenv}), it is easy to 
find that

\begin{equation}
\bar{n}(t)=\left[|\alpha(t)|^2+|\beta(t)|^2\right]\bar{n}
+|\beta(t)|^2,
\label{termt}
\end{equation}
where

\begin{equation}
\bar{n}=\frac{1}{{e}^{\beta\omega_0}-1}\;
\label{init}
\end{equation}
is the initial thermal equilibrium average for the number operator.

\subsection{The Time Independent Harmonic Oscillator}

Let us first consider the trivial case of the time
independent oscillator, with $\omega^2(t)=\omega_0^2$ as an
example of application.
In this case the solution for Eq. (\ref{oscil}) is just 

\begin{equation}
B(t)=ae^{i\omega_0 t}+be^{-i\omega_0 t}.
\label{sol1}
\end{equation}
substituting the above expression for $B(t)$ in Eq. (\ref{wrons1}) 
we obtain

\begin{equation}
|a|^2-|b|^2=\frac{1}{2\omega_0}.
\end{equation}
In this case we can obtain a time independent number operator choosing
$b=0$ and $|a|=1/\sqrt{2\omega_0}$. The other choice $a=0$ and $|b|\neq0$
is simply ruled out since in this case we would get $|b|^2<0$.
Choosing a null phase for $a$, we then obtain the solutions

\begin{equation}
\hat{a}(t)=\frac{{e}^{i\omega_0 t}}{\sqrt{2\omega_0}}
\left(\hat{p}-i\omega_0\hat{q}\right)
\label{opin1}
\end{equation}
and 

\begin{equation}
\hat{a}^\dag(t)=\frac{{e}^{-i\omega_0 t}}{\sqrt{2\omega_0}}
\left(\hat{p}+i\omega_0\hat{q}\right).
\label{opin2}
\end{equation}
Although the creation and annihilation operators are time dependent
the number operator $\hat{N}=\hat{a}^\dag(t)\hat{a}(t)$ is time
independent, as expected in this trivial example.

\subsection{An Exactly Soluble Time Dependent Harmonic Oscillator}

Let us consider now the case of an harmonic oscillator with the
time dependent frequency:

\begin{equation}
\omega^2(t)=\frac{\omega^2}{\cosh^2(t/\tau)}+\omega_0^2.
\label{toyt}
\end{equation}
Here the function $B(t)$ satisfies the differential equation

\begin{equation}
\ddot{B}(t)+\left[\frac{\omega^2}{\cosh^2(t/\tau)}+\omega_0^2\right]
B(t)=0.
\label{eqb}
\end{equation}
Eq. (\ref{eqb}) can be solved with the condition that at $t\rightarrow
-\infty$ the system is at thermal equilibrium at temperature $T$.
{}From the results of the previous subsection
we then have to solve Eq. (\ref{eqb}) with
the asymptotic condition that

\begin{equation}
B(t\rightarrow -\infty)=\frac{{e}^{i\omega_0t}}
{\sqrt{2\omega_0}}.
\label{asymt}
\end{equation}
The above problem is formally identical as an one-dimensional quantum
scattering problem. The solution of Eq. (\ref{eqb}), under the asymptotic
condition given by Eq. (\ref{asymt}) is \cite{solitons}

\begin{equation}
B(t)=\frac{2^{-i\tau\omega_0}}{\sqrt{2\omega_0}}\left[{\rm
sech}(t/\tau)\right]^{i\tau\omega_0} \;
{}F\left[\tilde{a},\tilde{b},\tilde{c},\frac{1}{2}+\frac{1}{2}
\tanh(t/\tau)
\right],
\label{solu2}
\end{equation}
where $F(x,y,z,w)$ is the Gauss' hypergeometric function
\cite{gradshteyn} and the coefficients $\tilde{a}$, $\tilde{b}$
and $\tilde{c}$ are given by

\begin{equation}
\tilde{a}=\frac{1}{2}+i\omega_0\tau+\sqrt{\omega^2\tau^2+\frac{1}{4}},
\label{co1}
\end{equation}

\begin{equation}
\tilde{b}=\frac{1}{2}+i\omega_0\tau-\sqrt{\omega^2\tau^2+\frac{1}{4}}
\label{co2}
\end{equation}
and

\begin{equation}
\tilde{c}=1+i\omega_0\tau.
\label{co3}
\end{equation}

When $t\rightarrow\infty$ intuitively one expects that the system
will thermalize. Taking the $t\rightarrow\infty$ limit in
Eq. (\ref{solu2}) we obtain the result

\begin{equation}
B(t\to\infty)=\frac{\Gamma(\tilde{c})\Gamma(\tilde{a}+\tilde{b}-\tilde{c})}
{\Gamma(\tilde{a})\Gamma(\tilde{b})}
\frac{{ e}^{i\omega_0
t}}{\sqrt{2\omega_0}}+\frac{\Gamma(\tilde{c})\Gamma(\tilde{c}-
\tilde{a}-\tilde{b})}{\Gamma(\tilde{c}-\tilde{a})\Gamma(\tilde{c}-
\tilde{b})}\frac{{ e}^{-i\omega_0 t}}{\sqrt{2\omega_0}}.
\label{asysol}
\end{equation}
Substituting Eq. (\ref{asysol}) in Eqs. (\ref{bug1}) and (\ref{bug2}) we
obtain for $\bar{n}(t)$, given by Eq. (\ref{termt}), an expression with
oscillatory behavior in general. {}For these cases clearly the system
never reaches thermal equilibrium. {}But if we analyze with some care
the arguments of the gamma-functions in Eq. (\ref{asysol}), we see that
for those values of $\tau$ such that $\tau^2\omega^2=N(N+1)$, with $N$
integer, then the second term in Eq. (\ref{asysol}) vanishes. In this
case the system reaches a final thermal equilibrium state with final
temperature that can be shown to be the same as of the initial one. This
same result was also obtained previously in Ref. \cite{jackiw} whose
authors have studied this problem solving the LvN equation obtained from 
a variational principle. We can also show that in the final thermal
equilibrium state there will be no dependences on parameters other than
those of the initial thermal equilibrium state ($\omega_0$ and $\beta$).
All the factors that appear multiplying the exponential factor in the
first term in Eq. (\ref{asysol}) contribute with a time independent
phase factor. This is equivalent to what happens in the one dimensional
scattering problem as consequence of the conservation of probability in
a reflectionless potential \cite{solitons}. In computing physical
quantities the phase factor cancels out and then, when the system reaches a
final thermal equilibrium state it loses all the information about how
this state was reached. Implicitly we have here a manifestation of
decoherence already at the level of this simple problem as a consequence
of a particular quench scenario and that will be so important when
models with larger degrees of freedom are taken into account. 

As an example of a concrete result, let us consider for convenience
the case $N=1$, {\it i.e,} when $\tau^2\omega^2=2$.
In this case $B(t)$ reduces to

\begin{equation}
B(t)=\frac{{ e}^{i\omega_0 t}}{\sqrt{2\omega_0}(1+i\omega_0\tau)}
\left[ i\omega_0\tau-\tanh(t/\tau)\right].
\label{toy1}
\end{equation}
Substituting  Eq. (\ref{toy1}) in Eqs. (\ref{bug1}) and (\ref{bug2}) we
get the results

\begin{equation}
|\alpha(t)|^2=1+\frac{\left[1-\tanh^2(t/\tau)\right]^2}
{4\omega_0^2\tau^2(1+\omega_0^2\tau^2)}
\label{alf1}
\end{equation}
and

\begin{equation}
|\beta(t)|^2= 
\frac{\left[1-\tanh^2(t/\tau)\right]^2}
{4\omega_0^2\tau^2(1+\omega_0^2\tau^2)}.
\label{alf2}
\end{equation}
Using these equations for $|\alpha(t)|^2$ and $|\beta(t)|^2$
in Eq. (\ref{termt}) we then obtain for $\bar{n}(t)$ the result:

\begin{equation}
\bar{n}(t)=\left(1+\frac{\left[1-\tanh^2(t/\tau)\right]^2}
{2\omega_0^2\tau^2(1+\omega_0^2\tau^2)}\right)\bar{n} +
\frac{\left[1-\tanh^2(t/\tau)\right]^2}
{4\omega_0^2\tau^2(1+\omega_0^2\tau^2)}.
\label{termt2}
\end{equation}

Results for $\bar{n}(t)$ for $\omega_0\beta=1$ and for different values 
of the "decay time" $\tau$ are shown in {}Fig. 1. Note that the peak at 
$t=0$ just represents the point around which the time-dependent frequency 
$\omega(t)$, Eq. (\ref{toyt}), reaches its maximum value and the peak is 
higher the 
smaller is the decay time, in agreement to the fact that, from 
Eq. (\ref{alf2}), the smallest is $\tau$ the largest is the production of 
excited states (particles, in the context of quantum fields),
as represented by $|\beta(t)|^2$. 

\section{Time Dependent Quantum Interacting Oscillators}

Lets us now consider a system of two time dependent oscillators,
linearly coupled with interaction strength $\lambda(t)$, 
with Hamiltonian given by
 
\begin{equation}
\hat{H}=\frac{1}{2}\sum_{j=1}^2\left[\hat{p}_j^2+
\omega_j^2(t)\hat{q}_j^2\right]
+\lambda(t)\hat{q}_1\hat{q}_2.
\label{couple}
\end{equation}
The time dependent annihilation and creation operators can now
be written as ($i=1,2$)

\begin{equation}
\hat{a}_i(t)=\sum_{j=1}^2 \left[
B_{ij}(t)\hat{p}_j-\dot{B}_{ij}(t)\hat{q}_j \right]
\label{2a}
\end{equation}
and

\begin{equation}
\hat{a}_i^\dag(t)=\sum_{j=1}^2 \left[
B_{ij}^\ast(t)\hat{p}_j-\dot{B}_{ij}^\ast(t)\hat{q}_j \right],
\label{2b}
\end{equation}
In an identical way 
as we have performed in the single time dependent oscillator of last
section, we require that the operators 
$\hat{a}_i(t)$ and $\hat{a}^\dag_i(t)$
satisfy the LvN equation, Eq. (\ref{lvn2}). 
{}From this requirement the following equations for the
time dependent coefficients $B_{ij}(t)$ follow:

\begin{equation}
\ddot{B}_{i1}(t)+\omega_1^2(t)B_{i1}(t)+\lambda(t)B_{i2}(t)=0
\label{2eqm1}
\end{equation}
and

\begin{equation}
\ddot{B}_{i2}(t)+\omega_2^2(t)B_{i2}(t)+\lambda(t)B_{i1}(t)=0.
\label{2eqm2}
\end{equation}

The condition

\begin{equation}
[\hat{a}_i(t),\hat{a}_j^\dag(t)]=\delta_{ij},
\label{alge2}
\end{equation}
leads now to the equation

\begin{equation}
\sum_{k=1}^2
\left[ 
\dot{B}_{ik}(t)B_{jk}^\ast(t)-B_{ik}(t)\dot{B}_{jk}^\ast(t)
\right]=
i\delta_{ij}
\label{wrons3}
\end{equation}
and the identities

\begin{equation}
[\hat{a}_i(t),\hat{a}_j(t)]=[\hat{a}_i^\dag(t),\hat{a}_j^\dag(t)]=0
\label{wrons31}
\end{equation}
lead, respectively, to the following additional equations 
for $B_{ij}(t)$:

\begin{eqnarray}
\sum_{k=1}^2
\left[ B_{ik}(t)\dot{B}_{jk}(t)-\dot{B}_{ik}(t)B_{jk}(t)
\right]&=&0,\nonumber\\
\sum_{k=1}^2
\left[
B_{ik}^\ast(t)\dot{B}_{jk}^\ast(t)-\dot{B}_{ik}^\ast(t)B_{jk}^\ast(t)
\right]&=&0.
\label{wrons32}
\end{eqnarray}

Using Eqs.  (\ref{2eqm1}) and (\ref{2eqm2}), it can be
shown that Eqs. (\ref{wrons3}), (\ref{wrons31}) and
(\ref{wrons32}) must be time independent. As in the one single
harmonic oscillator case we can write the time dependent
annihilation and creation operators as functions of the initial
annihilation and creation operators (that we just call by 
$\hat{a}_i$ and
$\hat{a}_i^\dag$, respectively, without the time argument,
as indicating that these quantities should be evaluated at
the initial time). This can be accomplished by using the 
requirement
that initially the system consists of two independent harmonic
oscillators with frequencies $\omega_1$ and $\omega_2$,
each one at well defined thermal equilibrium states with temperatures 
$1/\beta_1$ and $1/\beta_2$, respectively. In this case the 
matrix $\{B\}$, at the initial time, has elements $B_{ij}$ given by

\begin{equation}
B_{ij}=\delta_{ij} \frac{e^{i\omega_i t}}{\sqrt{2\omega_i}}.
\label{initb}
\end{equation}
We can now readily show that the Bogoliubov transformation in this
case is given by

\begin{equation}
\left(\begin{array}{c}\hat{a}_i(t)\\
\hat{a}_i^\dag(t)\end{array}\right)=
\sum_{j=1}^2
\left( \begin{array}{cc}\alpha_{ij}(t)&\beta_{ij}(t)\\
\beta_{ij}^\ast(t)&\alpha_{ij}^\ast(t)\end{array}\right)
\left(\begin{array}{c}\hat{a}_j\\
\hat{a}_j^\dag\end{array}\right),
\label{bugo2}
\end{equation}
where
\begin{equation}
\alpha_{ij}(t)=i\left(\dot{B}_{jj}^\ast B_{ij}(t)-B_{jj}^\ast 
\dot{B}_{ij}(t)
\right)
\label{bug21}
\end{equation}
and 

\begin{equation}
\beta_{ij}(t)=i\left(B_{jj}\dot{B}_{ij}(t)-
\dot{B}_{jj}B_{ij}(t)\right).
\label{bug22}
\end{equation}

In terms of the above Bogoliubov coefficients the
conditions (\ref{wrons3}) and (\ref{wrons32}) are given,
respectively, by

\begin{equation}
\sum_{k=1}^2
\left[
\alpha_{ik}(t)\alpha_{jk}^\ast(t)-\beta_{ik}(t)\beta_{jk}^\ast(t)
\right]=
\delta_{ij}
\label{wrobu}
\end{equation}
and

\begin{eqnarray}
\sum_{k=1}^2
\left[
\alpha_{ik}(t)\beta_{jk}(t)-\beta_{ik}(t)\alpha_{jk}(t)
\right]&=&0,\nonumber\\
\sum_{k=1}^2
\left[
\alpha_{ik}^\ast(t)\beta_{jk}^\ast(t)-\beta_{ik}^\ast(t)
\alpha_{jk}^\ast(t)
\right]&=&0.
\label{wrobu1}
\end{eqnarray}
Since $\hat{a}_i(t)$ and $\hat{a}_i^\dag(t)$ are solutions of the 
LvN equation, the exact solution for the time dependent density
operator with thermal equilibrium initial conditions is

\begin{equation}
\hat{\rho}(t)=\frac{1}{Z_{\beta_1}Z_{\beta_2}}
\exp\left\{-\sum_{i=1}^{2}\beta_i\omega_i
\left[\hat{a}_i^\dag(t)\hat{a}_i(t)+\frac{1}{2}\right]\right\},
\label{sold2}
\end{equation}
where $Z_{\beta_i}$, $i=1,2$, are the initial (equilibrium) 
partition functions for each
oscillator as given by Eq. (\ref{parteq}).
Note that the out-of-equilibrium problem resulting from the coupling of 
the two systems at an initial time $t_0$ is reduced to solving the
two coupled second order differential equations given by
Eqs. (\ref{2eqm1}) and (\ref{2eqm2}). As in the time dependent single
harmonic oscillator case we can compute the time
dependent thermal expectation values for the number operators,
$\hat{N}_i=\hat{a}_i^\dag\hat{a}_i$, 

\begin{eqnarray}
\bar{n}_i(t)&=&{\rm Tr}\left[\hat{\rho}(t)
\hat{a}_i^\dag\hat{a}_i\right] \nonumber\\
&=&\frac{1}{Z_{\beta_1} Z_{\beta_2}}\sum_{n_1,n_2=0}^{\infty}
\langle n_1(t),n_2(t) |
\exp\left\{-\sum_{j=1}^{2}\beta_j\omega_j
\left[\hat{a}_j^\dag(t)\hat{a}_j(t)+\frac{1}{2}\right]\right\}\nonumber\\
& &~~~~~~~~~~~~~~~~~~~~~~~~~~~~~~~~~~~~~~~~
\hat{a}_i^\dag\hat{a}_i| n_1(t),n_2(t)\rangle .
\label{solvw1}
\end{eqnarray}
In order to evaluate the above expression we have to solve
Eq. (\ref{bugo2}) for $\hat{a}_i$ and $\hat{a}_j^\dag$ in terms of
$\hat{a}_i(t)$, $\hat{a}_j^\dag(t)$. This can be easily done using 
Eqs. (\ref{wrobu}) and (\ref{wrobu1}) then obtaining

\begin{equation}
\left(\begin{array}{c}\hat{a}_i\\
\hat{a}_i^\dag\end{array}\right)=
\sum_{j=1}^2
\left( \begin{array}{cc}\alpha_{ij}^\ast(t)&-\beta_{ji}(t)\\
-\beta_{ji}^\ast(t)&\alpha_{ij}(t)\end{array}\right)
\left(\begin{array}{c}\hat{a}_j(t)\\
\hat{a}_j^\dag(t)\end{array}\right).
\label{bugo23}
\end{equation}
Using  Eq. (\ref{bugo23}) in Eq. (\ref{solvw1}) we obtain

\begin{equation}
\bar{n}_i(t)=\sum_{j=1}^2\left[(|\alpha_{ij}(t)|^2+
|\beta_{ji}(t)|^2)
\bar{n}_j+|\beta_{ji}(t)|^2\right],
\label{pro2}
\end{equation}
which generalizes Eq. (\ref{termt}) for the case of more than
one quantum oscillator.  
In Eq. (\ref{pro2}) $\bar{n}_j$ is the initial thermal equilibrium
distribution,

\begin{equation}
\bar{n}_j=\frac{1}{{e}^{\beta_j\omega_j}-1}.
\label{init2o}
\end{equation}

\subsection{Solution for the Instantaneous Interaction}

As a first application of the above equations, let us consider 
the example of two coupled quantum oscillators
with time independent frequencies and coupling 
$\lambda(t)$ suddenly turned on at time $t=0$, {\it i.e},

\begin{equation}
\lambda(t)=\left\{\begin{array}{c}0,~ t\leq 0\\\lambda,~t>0\end{array}
\right..
\label{sudla}
\end{equation}
In this case  we have for $t\leq 0$ that

\begin{equation}
B_{ij}(t\leq 0)=\left(\begin{array}{cc}\frac{e^{i\omega_1
t}}{\sqrt{2\omega_1}}&0\\0&\frac{e^{i\omega_2
t}}{\sqrt{2\omega_2}}\\\end{array}\right)
\label{tmen}
\end{equation}
and for $t>0$ we can  solve Eqs. (\ref{2eqm1}) and (\ref{2eqm2})
in terms of normal coordinates $Q_{ij}(t)$, 
defined in terms of $B_{ij}(t)$ by

\begin{equation}
B_{ij}(t)=\sum_{k=1}^2 \eta_{jk}Q_{ik}(t),
\label{a1}
\end{equation}
where the matrix transformation $\eta_{ij}$ is time independent.
Substituting the above equation in Eqs. (\ref{2eqm1}) and (\ref{2eqm2}) 
we obtain (for details see appendix A)

\begin{equation}
\ddot{Q}_{ik}(t)+\Omega_k^2Q_{ik}(t)=0,
\label{a2}
\end{equation}
provided that $\eta_{ij}$ is an orthogonal matrix with elements 
given by

\begin{equation}
\eta_{1j}=\left[1+\frac{\lambda^2}{(\omega_2^2-\Omega_j^2)^2}
\right]^{-\frac{1}{2}},
\label{a3}
\end{equation}

\begin{equation}
\eta_{2j}=\frac{\lambda}{\Omega_j^2-\omega_2^2}\eta_{1j},
\label{a4}
\end{equation}
and $\Omega_j$ are obtained as the solutions of

\begin{equation}
(\omega_1^2-\Omega_j^2)(\omega_2^2-\Omega_j^2)=\lambda^2,
\label{a5}
\end{equation}
from which we obtain

\begin{eqnarray}
&&\Omega_1^2=\frac{1}{2}\left[\omega_1^2+\omega_2^2+\sqrt{(\omega_1^2-
\omega_2^2)^2+4\lambda^2}~\right],
\nonumber \\
&& \Omega_2^2=\frac{1}{2}\left[\omega_1^2+\omega_2^2-\sqrt{(\omega_1^2-
\omega_2^2)^2+4\lambda^2}~\right].
\label{a6}
\end{eqnarray}

Solving Eq. (\ref{a2}) and substituting it in Eq. (\ref{a1}),
we obtain for $B_{ij}(t)$ the following expression

\begin{equation}
B_{ij}(t)=\sum_{k=1}^2 \eta_{jk}\left(a_{ik}{ e}^{i\Omega_kt}+
b_{ik}{ e}^{-i\Omega_kt}\right),
\label{a7}
\end{equation}
where $a_{ik}$ and $b_{ik}$ are constants of integration to be 
determined by the initial conditions. Since at $t=0$, $B_{ij}(t)$
and $\dot{B}_{ij}(t)$ are continuous, then from Eqs. (\ref{tmen})
and (\ref{a7}), we obtain that the coefficients $a_{ik}$ and $b_{ik}$,
appearing in Eq. (\ref{a7}), are given by

\begin{eqnarray}
&&a_{ik}=\frac{\eta_{ik}}{\sqrt{8\omega_i}}\left(1+\frac{\omega_i}
{\Omega_k}\right), \nonumber \\
&& b_{ik}=\frac{\eta_{ik}}{\sqrt{8\omega_i}}\left(1-\frac{\omega_i}
{\Omega_k}\right).
\label{a8}
\end{eqnarray}
Substituting Eq. (\ref{a8}) back in Eq. (\ref{a7}), we obtain
the result

\begin{equation}
B_{ij}(t)=\sum_{k=1}^2 \frac{\eta_{ik}\eta_{jk}}{\sqrt{2\omega_i}}
\left[\cos(\Omega_kt)
+\frac{i\omega_i}{\Omega_k}\sin(\Omega_kt)\right]
\label{a9}
\end{equation}
Using Eq. (\ref{tmen}) for  $B_{ij}$ at $t=0$
in Eq. (\ref{bug21}) we obtain for the Bogoliubov coefficients the
expressions

\begin{equation}
\alpha_{ij}(t)=\sqrt{\frac{\omega_j}{2}}B_{ij}(t)-\frac{i}
{\sqrt{2\omega_j}}\dot{B}_{ij}(t)
\label{solfa1}
\end{equation}
and

\begin{equation}
\beta_{ij}(t)=\sqrt{\frac{\omega_j}{2}}B_{ij}(t)+\frac{i}
{\sqrt{2\omega_j}}\dot{B}_{ij}(t).
\label{solfa2}
\end{equation}

Eqs. (\ref{a9}) - (\ref{solfa2}) together with Eqs. 
(\ref{a3}) - (\ref{a6})
completely solve the out-of-equilibrium evolution of the 
system in consideration. In this case the system never reaches
a final thermal equilibrium state since the 
Bogoliubov coefficients obtained using the result for $B_{ij}(t)$,
as given by Eq. (\ref{a9}), have an oscillatory behavior for all $t$.
The solutions for $\bar{n}_1(t)$ and $\bar{n}_2 (t)$ are shown
in {}Fig. 2 for the case of $\omega_1=\omega_2$. This will be further
discussed below.

\subsection{An Exactly Soluble Smooth Quench}

As an example of non-instantaneous quench and exactly soluble
model, let us consider next the case of a smooth quench with
the linear coupling given by the expression

\begin{equation}
\lambda(t)=\frac{\lambda}{2}\left[1+\tanh(t/\tau)\right],
\label{ltem}
\end{equation}
and with time independent frequencies $\omega_1=\omega_2=\omega$.
The equations for the $B_{ij}(t)$ functions now become 

\begin{equation}
\ddot{B}_{i1}(t)+\omega^2B_{i1}(t)+\frac{\lambda}{2}
[1+\tanh(t/\tau)]B_{i2}(t)=0
\label{eqmv1}
\end{equation}
and

\begin{equation}
\ddot{B}_{i2}(t)+\omega^2B_{i2}(t)+\frac{\lambda}{2}
[1+\tanh(t/\tau)]B_{i1}(t)=0.
\label{eqmv2}
\end{equation}
Eqs. (\ref{eqmv1}) and (\ref{eqmv2}) can be more
easily solved if we make use of the following transformation:

\begin{eqnarray}
R_i(t)&=&B_{i1}(t)-B_{i2}(t)\nonumber\\
G_i(t)&=&B_{i1}(t)+B_{i2}(t),
\label{combin}
\end{eqnarray}
which then lead to the following uncoupled differential 
equations for $R_i(t)$ and $G_i(t)$, 

\begin{equation}
\ddot{R}_i(t)+\left\{\omega^2-\frac{\lambda}{2}[1+\tanh(t/\tau)]
\right\}R_i(t)=0
\label{eqr1}
\end{equation}
and

\begin{equation}
\ddot{G}_i(t)+\left\{\omega^2+\frac{\lambda}{2}[1+\tanh(t/\tau)]
\right\}G_i(t)=0.
\label{eqr2}
\end{equation}
Note that even for the case of time dependent frequencies
$\omega_i(t)$, the Eqs. (\ref{eqmv1}) and (\ref{eqmv2}) still uncouple
and can be put in the form of Eqs. (\ref{eqr1}) and (\ref{eqr2}).
The solutions for Eqs. (\ref{eqr1}) and (\ref{eqr2}) can now be
found to be given by

\begin{equation}
R_i(t) =A_i { e}^{-i\omega t}F[-a_1,-b_1,c^\ast,-{e}^{2t/\tau}] 
+ B_i { e}^{i\omega t}F[b_1,a_1,c,-{e}^{2t/\tau}]
\label{solRi}
\end{equation}
and

\begin{equation}
G_i(t) =C_i { e}^{-i\omega t}F[-a_2,-b_2,c^\ast,-{e}^{2t/\tau}] 
+D_i { e}^{i\omega t}F[b_2,a_2,c,-{e}^{2t/\tau}],
\label{solGi}
\end{equation}
where

\begin{eqnarray}
&&a_1=\frac{i \tau}{2} (\omega - \Omega_2),\nonumber \\
&&a_2=\frac{i \tau}{2} (\omega - \Omega_1),\nonumber\\
&&b_1=\frac{i \tau}{2} (\omega + \Omega_2),\nonumber \\
&&b_2=\frac{i \tau}{2} (\omega + \Omega_1),\nonumber \\
&&c=1+i\omega\tau,\nonumber\\
\end{eqnarray}
and $\Omega_{1(2)} = \sqrt{\omega^2 \pm |\lambda|}$. We assume that 
$\omega^2 > |\lambda|$ in order to avoid runaway solutions.
Inverting the transformation given by (\ref{combin}) we now obtain
that

\begin{eqnarray}
B_{i1} &=& \frac{1}{2} \left\{ C_i { e}^{-i\omega t}
{}F[-a_2,-b_2,c^\ast,-{e}^{2t/\tau}] +
D_i { e}^{i\omega t}
{}F[b_2,a_2,c,-{e}^{2t/\tau}] 
\right. \nonumber \\
&& ~~+\left.  A_i { e}^{-i\omega t} 
{}F[-a_1,-b_1,c^\ast,-{e}^{2t/\tau}] +
B_i { e}^{i\omega t}
{}F[b_1,a_1,c,-{e}^{2t/\tau}]
\right\},
\end{eqnarray}
and

\begin{eqnarray}
B_{i2} &=& \frac{1}{2} \left\{ C_i { e}^{-i\omega t}
{}F[-a_2,-b_2,c^\ast,-{e}^{2t/\tau}] +
D_i { e}^{i\omega t}
{}F[b_2,a_2,c,-{e}^{2t/\tau}] 
\right. \nonumber \\
&& ~~-\left.  A_i { e}^{-i\omega t} 
{}F[-a_1,-b_1,c^\ast,-{e}^{2t/\tau}] -
B_i { e}^{i\omega t}
{}F[b_1,a_1,c,-{e}^{2t/\tau}]
\right\},
\end{eqnarray}
Using the initial conditions (at $t\to - \infty$),

\begin{eqnarray}
&&B_{11} = B_{22} = \frac{\exp(i \omega t)}{\sqrt{2 \omega}}\nonumber \\
&&B_{12} = B_{21} =0,
\end{eqnarray}
we obtain the constants $A_i,B_i, C_i$ and $D_i$ as
given by:
$A_1=A_2=C_1=C_2=0$ and $B_1=-B_2=D_1=D_2 = 1/\sqrt{2 \omega}$. Then,
finally, we find the solutions for the time dependent factors
$B_{ij}(t)$ as given by the expressions

\begin{eqnarray}
&&B_{11} (t)= B_{22} (t)= \frac{{ e}^{i \omega t}}
{2 \sqrt{2 \omega}}\left\{
{}F[b_1,a_1,c,-{e}^{2t/\tau}]
+F[b_2,a_2,c, -{e}^{2t/\tau}]
\right\},\nonumber\\
&&B_{12} (t)=B_{21} (t) =\frac{{ e}^{i \omega t}}
{2 \sqrt{2 \omega}}\left\{
{}F[b_1,a_1,c, -{e}^{2t/\tau}]
-F[b_2,a_2,c, -{e}^{2t/\tau}]
\right\}.
\end{eqnarray}
The asymptotic behavior ($t \gg \tau$) for the functions $B_{ij}(t)$ 
follows from the asymptotic behavior of the hypergeometric 
functions \cite{abramov}, which leads to the results

\begin{eqnarray}
B_{11}(t\to\infty)&=&\frac{1}{\sqrt{8\omega_0}}
\left[\frac{\Gamma(c)\Gamma(a_1-b_1)}
{\Gamma(a_1)\Gamma(c-b_1)}\; { e}^{i \Omega_2 t}+
\frac{\Gamma(c)\Gamma(b_1-a_1)}
{\Gamma(b_1)\Gamma(c-a_1)}\; { e}^{-i \Omega_2 t}
\right.\nonumber\\
& &~~~~~~+\left.\frac{\Gamma(c)\Gamma(a_2-b_2)}
{\Gamma(a_2)\Gamma(c-b_2)}\; { e}^{i \Omega_1 t}+
\frac{\Gamma(c)\Gamma(b_2-a_2)}
{\Gamma(b_2)\Gamma(c-a_2)}\; { e}^{-i \Omega_1 t}
\right]
\label{assim1}
\end{eqnarray}
and

\begin{eqnarray}
B_{12}(t\to\infty)&=&\frac{1}{\sqrt{8\omega_0}}
\left[\frac{\Gamma(c)\Gamma(a_1-b_1)}
{\Gamma(a_1)\Gamma(c-b_1)}\; { e}^{i \Omega_2 t}+
\frac{\Gamma(c)\Gamma(b_1-a_1)}
{\Gamma(b_1)\Gamma(c-a_1)}\; 
{ e}^{-i \Omega_2 t}\right.\nonumber\\
& &~~~~~~-\left.\frac{\Gamma(c)\Gamma(a_2-b_2)}
{\Gamma(a_2)\Gamma(c-b_2)}\; { e}^{i \Omega_1 t}-
\frac{\Gamma(c)\Gamma(b_2-a_2)}
{\Gamma(b_2)\Gamma(c-a_2)}\; { e}^{-i \Omega_1 t}
\right].
\label{assim2}
\end{eqnarray}

Note that Eqs. (\ref{assim1}) and (\ref{assim2}) are similar to
Eq. (\ref{a7}) for the case of an instantaneous quench and with 
$\omega_1=\omega_2$. Therefore we expect that both cases will 
lead to qualitative similar results in the asymptotic limit
with both oscillators never reaching a final equilibrium state. 
This is a consequence that in both cases the models considered are 
purely reversible systems and therefore energy is transferred
in a reversible way from one oscillator to the other, with the system
never actually equilibrating or thermalizing. This situation
is changed considerably when the system is coupled to an
environment, modeled for instance as infinity other oscillators
coupled to one given oscillator, just like in the
Caldeira-Leggett type of models \cite{caldeira}.
Under this situation energy can now be transferred irreversibly
from the system of oscillators to the environment leading 
eventually to equilibration and thermalization of the system.
How these processes come about will be explicitly shown in the next 
section in the context of the LvN approach.

\section{A system in interaction with a bath: an oscillator interacting with 
$N\to\infty$ other oscillators}

In this section we consider a system composed by an oscillator
interacting linearly with $N\to\infty$ other oscillators
(here to be considered as the environment or bath, in the typical
denomination of Caldeira-Leggett type of models \cite{caldeira}). 
The Hamiltonian  is now described by 

\begin{equation}
\hat{H}=\frac{1}{2}\hat{p}_0^2+\frac{\omega_0^2}{2}\hat{q}_0^2+
\frac{1}{2}\sum_{k=1}^N
\left[\hat{p}_k^2+\omega_k^2 \hat{q}_k^2+2\lambda_k(t)\hat{q}_0
\hat{q}_k\right].
\label{i1}
\end{equation}
In the above $\lambda_k(t)$ is a time dependent coupling constant (with 
dimension of the square of a frequency). In general the frequencies can 
also be made time
dependent, but this case will not be explicitly considered here. 
 
We consider that initially the oscillator and the environment are each 
separately at thermal equilibrium states at inverse temperatures $\beta_0$
and $\beta$, respectively (we do not concern here about how this initial
state can be realized physically). That is, we consider that initially 
$\lambda_k(t)=0$.
Then the initial density operator $\hat{\rho}_0$, of the total system 
will be given by 

\begin{equation}
\hat{\rho}_0=\hat{\rho}_{\beta_0}\otimes\hat{\rho}_{\beta},
\label{i2}
\end{equation}
where $\hat{\rho}_{\beta_0}$ and $\hat{\rho}_{\beta}$ are the initial
thermal equilibrium density operators. We want to investigate how the
initial thermal equilibrium state for the oscillator evolves in time
when it is coupled to the environment. More explicitly, we want to study 
the equilibration process and in particular we would like to know how
is the final equilibrium distribution reached and how long it takes to
reach that state. 

In order to solve for the
density operator in an exact way we proceed in an identical way 
as in last section. It is clear that analogous expressions to
those obtained in the last section will also be obtained here,
with similar arguments from the LvN approach being carried out to this
problem here. 
{}For example, the thermal expectation value of the number operator can be
shown that can be put in the form, in terms of Bogoliubov
coefficients, as

\begin{equation}
\bar{n}_\mu(t)=\sum_{\nu=0}^N\left[\left(|\alpha_{\mu\nu}(t)|^2+
|\beta_{\nu\mu}(t)|^2\right)\bar{n}_\nu+|\beta_{\nu\mu}(t)|^2\right],
\label{i3}
\end{equation}
where $\mu=(0,k)$, $k=1,2..$ and $\bar{n}_0$, $\bar{n}_k$ are the initial 
thermal equilibrium distributions for the oscillator and the environment,
respectively,  given by

\begin{equation}
\bar{n}_0=\frac{1}{{ e}^{\beta_0\tilde{\omega}}-1},
\label{i4}
\end{equation}
and

\begin{equation}
\bar{n}_k=\frac{1}{{ e}^{\beta\omega_k}-1}.
\label{ii4}
\end{equation}
Note that in Eq. (\ref{i4}) we have used $\tilde{\omega}$ instead of
$\omega_0$. As will become clearer in the derivation below, $\omega_0$
will represent a bare frequency, while zero point corrections coming from
the coupling to the (infinity) environment oscillators leads to the
definition of a physical (renormalized) frequency, that here we
denote it by $\tilde{\omega}$. The Bogoliubov
coefficients are given by equations analogous to Eqs. 
(\ref{bug21})-(\ref{bug22})

\begin{equation}
\alpha_{\mu\nu}(t)=i\left(\dot{B}_{\nu\nu}^\ast B_{\mu\nu}(t)-
B_{\nu\nu}^\ast\dot{B}_{\mu\nu}(t)\right),
\label{i5}
\end{equation}
\begin{equation}
\beta_{\mu\nu}(t)=i\left(B_{\nu\nu} \dot{B}_{\mu\nu}(t)-\dot{B}_{\nu\nu}
B_{\mu\nu}(t)\right),
\label{i6}
\end{equation}
where $B_{\mu\nu}$ are the non zero initial values of $B_{\mu\nu}(t)$,

\begin{equation}
B_{\mu\nu}=\delta_{\mu\nu}\frac{{ e}^{i\tilde{\omega}_\mu t}}
{\sqrt{2\tilde{\omega}_\mu}},
\label{i7}
\end{equation}
where $\tilde{\omega}_\mu=(\tilde{\omega}, \omega_k)$. 
Of course, the Bogoliubov coefficients also follow identical relations
as given by Eqs. (\ref{wrobu})-(\ref{wrobu1}). The coefficients 
$B_{\mu\nu}(t)$ also obey equations similar to Eqs. 
(\ref{2eqm1})-(\ref{2eqm2}),

\begin{equation}
\ddot{B}_{\mu 0}(t)+\omega_0^2B_{\mu 0}(t)+
\sum_{k=1}^N\lambda_k(t)B_{\mu k}(t)=0,
\label{i8}
\end{equation}
\begin{equation}
\ddot{B}_{\mu k}(t)+\omega_k^2B_{\mu k}(t)+\lambda_k(t)B_{\mu 0}(t)=0.
\label{i9}
\end{equation}
In general Eqs. (\ref{i8})-(\ref{i9}) cannot be solved analytically
for any arbitrary time dependence of $\lambda(t)$. There is a situation 
however in 
which these equations can be solved exactly, and this is the case when 
$\lambda(t)$ is suddenly turned on at some given time, for example at $t=0$.
{}For convenience we can then express $\lambda_k(t)$ as

\begin{equation}
\lambda_k(t)=\left\{\begin{array}{lc}0, & ~ t\leq 0\\ 
\omega_k\sqrt{g\Delta \omega}, &
~t>0\end{array}
\right..
\label{add1}
\end{equation}
where $g$ is now a constant (coupling constant) parameter introduced
for convenience and $\Delta \omega$ is defined below. 
Also in order to  simplify our calculations we start with $\omega_k$ 
given by

\begin{equation}
\omega_k=\frac{k\pi}{L}, ~k=1,2,\ldots
\label{discreto}
\end{equation}
and then we go to the continuum limit at the end by letting $L\to\infty$. 
In Eq. (\ref{add1}) $\Delta \omega=\frac{\pi}{L}$ is the interval between 
two neighboring frequencies.
For $t<0$ the solution of the system of Eqs. (\ref{i8})-(\ref{i9}) is given 
by Eq. (\ref{i7}). For $t>0$ the solution is found by introducing,
as in Sec. III,  normal 
coordinates $Q_{\mu\nu}(t)$ now defined by 

\begin{equation}
B_{\mu\nu}(t)=\sum_{\rho=0}^N\eta_{\nu\rho}Q_{\mu\rho}(t),
\label{i12}
\end{equation}
where $Q_{\mu\rho}(t)$ satisfy
\begin{equation}
\ddot{Q}_{\mu\rho}(t)+\Omega_\rho^2Q_{\mu\rho}(t)=0,
\label{add2}
\end{equation}
and $\eta_{\mu\nu}$ is an orthogonal matrix with elements (for details 
see appendix A) given by 

\begin{equation}
\eta_{k\mu}=\frac{\omega_k\sqrt{g\Delta \omega}}{\Omega_\mu^2-
\omega_k^2}\eta_{0\mu},
\label{i14}
\end{equation}
\begin{equation}
\eta_{0\mu}=\left[1+\sum_{k=1}^N\frac{g\omega_k^2\Delta \omega}
{(\omega_k^2-\Omega_\nu^2)^2}\right]^{-\frac{1}{2}},
\label{i15}
\end{equation}
with  normal frequencies $\Omega_\mu$ obtained as the solutions of

\begin{equation}
\omega_0^2-\Omega_\mu^2=\sum_{k=1}^{N}\frac{g\omega_k^2\Delta \omega}
{\omega_k^2-\Omega_\mu^2}.
\label{i16}
\end{equation}
Eq. (\ref{i16}) can be rewritten as

\begin{equation}
\omega_0^2-gN\Delta\omega-\Omega_\mu^2=\sum_{k=1}^{N}
\frac{g\Omega_\mu^2\Delta \omega}{\omega_k^2-\Omega_\mu^2}.
\label{i17}
\end{equation}

By taking the limit $N\to\infty$ in Eq. (\ref{i17}) we see that
the above equation make no sense, since in such case the left hand side
is infinity while the right hand is not. This is a consequence of
the (infinity) zero point energy added as we let $N\to \infty$.
In order to overcome this
problem we introduce a renormalized frequency $\tilde{\omega}$
defined as

\begin{equation}
\tilde{\omega}=\sqrt{\omega_0^2-gN\Delta\omega}.
\label{i18}
\end{equation}
In order to $\tilde{\omega}$ be finite we see that $\omega_0^2$ have 
to be infinity. Then we conclude that $\omega_0$ is not the physical 
frequency of the oscillator, it is a bare (unrenormalized) frequency. 
The physical frequency is the finite renormalized frequency 
$\tilde{\omega}$. Also the introduction of the renormalized frequency
guarantees that all the solutions $\Omega^2$ of Eq. (\ref{i17})
be positive allowing us to avoid possible runaway solutions 
\cite{adolfo}.

Using Eq. (\ref{discreto}) and taking $N,~L\to\infty$ in Eqs. 
(\ref{i14})-(\ref{i16}) we obtain

\begin{equation}
\eta_{0\mu}=\lim_{\Delta \omega\to 0}\frac{2\Omega_\mu
\sqrt{g\Delta\omega}}{\sqrt{4(\Omega_\mu^2-\tilde{\omega}^2)^2+
\pi^2g^2\Omega_\mu^2}},
\label{i19}
\end{equation}
\begin{equation}
\eta_{k\mu}=\frac{2g\omega_k\Delta\omega}{(\Omega_\mu^2-\omega_k^2)}
\frac{\Omega_\mu}{\sqrt{4(\Omega_\mu^2-\tilde{\omega}^2)^2+
\pi^2g^2\Omega_\mu^2}}
\label{i20}
\end{equation}
and 

\begin{equation}
{\rm cot}(L\Omega)=\frac{2\Omega}{\pi g}+\frac{\Delta \omega}
{\pi \Omega}\left(1-\frac{2\tilde{\omega}^2}{g\Delta\omega}\right),
\label{i21}
\end{equation}
where in Eq. (\ref{i21}) we have dropped the index $\mu$ for simplicity 
and it is to be
understood  
that $\Omega_\mu$ are the roots of Eq. (\ref{i21}). 

Substituting  the solutions of Eq. (\ref{add2}) in Eq. (\ref{i12}) leads to 
the following expression for $B_{\mu \nu}(t)$, 

\begin{equation}
B_{\mu\nu}(t)=\sum_{\rho=0}^N\eta_{\nu\rho}\left(a_{\mu\rho}
e^{i\Omega_\rho t}+b_{\mu\rho}e^{-i\Omega_\rho t}\right).
\label{ad4}
\end{equation}
{}From the continuity of $B_{\mu\nu}(t)$ and of its first derivative
at $t=0$ we can obtain $a_{\mu\rho}$ and $b_{\mu\rho}$ (see appendix A).
Substituting the obtained expressions for these coefficients in Eq. 
(\ref{ad4}) we obtain the result

\begin{equation}
B_{\mu\nu}(t)=\sum_{\rho=0}^N
\frac{\eta_{\mu\rho}\eta_{\nu\rho}}{\sqrt{8\tilde{\omega}_\mu}}
\left[\left(1+\frac{\tilde{\omega}_\mu}{\Omega_\rho}\right)
{ e}^{i\Omega_\rho t}
+\left(1-\frac{\tilde{\omega}_\mu}{\Omega_\rho}\right)
{ e}^{-i\Omega_\rho t}\right],
\label{i22}
\end{equation}
from which the Bogoliubov coefficients follow

\begin{equation}
\alpha_{\mu\nu}(t)=\sqrt{\frac{\tilde{\omega}_\nu}{2}}
B_{\mu\nu}(t)-\frac{i}{\sqrt{2\tilde{\omega}_\nu}}\dot{B}_{\mu\nu}(t)
\label{i23}
\end{equation}
and

\begin{equation}
\beta_{\mu\nu}(t)=\sqrt{\frac{\tilde{\omega}_\nu}{2}}
B_{\mu\nu}(t)+\frac{i}{\sqrt{2\tilde{\omega}_\nu}}\dot{B}_{\mu\nu}(t).
\label{i24}
\end{equation}
Taking $\mu=0$ in Eq. (\ref{i3}), we obtain the time dependent
thermal expectation value of the number operator for the oscillator
$q_0$, 

\begin{eqnarray}
\bar{n}_0(t)&=&[\left(|\alpha_{00}(t)|^2+
|\beta_{00}(t)|^2\right)\bar{n}_0+|\beta_{00}(t)|^2\nonumber\\
& &~~~~~~~~~~~~~~
+\sum_{k}\left[\left(|\alpha_{0 k}(t)|^2
+|\beta_{k 0}(t)|^2\right)\bar{n}_k+|\beta_{k 0}(t)|^2\right].
\label{i25}
\end{eqnarray}
Then in order to obtain $\bar{n}_0(t)$ we have first to compute
$\alpha_{00}(t)$, $\alpha_{0k}(t)$, $\beta_{00}(t)$ and 
$\beta_{k0}(t)$. These coefficients are obtained in Appendix B.
They are given by Eqs. (\ref{B6}), (\ref{B7}), (\ref{B14}) and 
(\ref{B15}). {}From Eqs. (\ref{B6})-(\ref{B7}) we can see that

\begin{equation}
\alpha_{00}(t\to\infty)=\beta_{00}(t\to\infty)=0.
\label{i26}
\end{equation}
On the other hand from Eqs. (\ref{B14})-(\ref{B15}) we obtain

\begin{equation}
\alpha_{0k}(t\to\infty)=\sqrt{\frac{\omega_k}{\tilde{\omega}}}
\frac{(\tilde{\omega}+\omega_k)(2\omega_k^2-2\tilde{\omega}^2+i\pi g
\omega_k)}{\left[4\left(\omega_k^2-\tilde{\omega}^2\right)^2+
\pi^2g^2\omega_k^2\right]}
\sqrt{g\Delta \omega}e^{i\omega_k t},
\label{app14}
\end{equation}
\begin{equation}
\beta_{k0}(t\to\infty)=\sqrt{\frac{\omega_k}{\tilde{\omega}}}
\frac{(\tilde{\omega}-\omega_k)(2\omega_k^2-2\tilde{\omega}^2+i\pi g
\omega_k)}{\left[4\left(\omega_k^2-\tilde{\omega}^2\right)^2+
\pi^2g^2\omega_k^2\right]}
\sqrt{g\Delta \omega}e^{i\omega_k t}.
\label{app15}
\end{equation}
Now substituting the above equations in Eq. (\ref{i25}) and going to the
continuum limit we obtain for $\bar{n}_0(t\to\infty)$ the following
result

\begin{eqnarray}
\bar{n}_0(t\to\infty)&=&\frac{2g}{\tilde{\omega}}
\int_0^\infty d\omega\frac{\omega(\omega^2+\tilde{\omega}^2)}
{\left[4(\omega^2-\tilde{\omega}^2)^2+\pi^2g^2\omega^2\right]}
\frac{1}{\left(e^{\beta\omega}-1\right)}\nonumber\\
& &~~~~~~~~~~~~~~~~~~+\frac{g}{\tilde{\omega}}\int_0^\infty d \omega
\frac{\omega(\omega-\tilde{\omega})^2}{4(\omega^2-\tilde{\omega}^2)^2
+\pi^2g^2\omega^2}.
\label{i28}
\end{eqnarray}
Note that, as a consequence of Eq. (\ref{i26}), 
the asymptotic result for the number density of the system's
oscillator $q_0$ does not depend on its initial temperature, only on
the (continuum) ensemble initial temperature $T=1/\beta$.
We also note that while the first term on the right-hand-side in Eq. 
(\ref{i28}) is convergent, the second term diverges logarithmically for 
$\omega\to\infty$.
It is easy to prove that this divergent term is equal to the
asymptotic value of $\langle 0|\hat{a}^\dag_0(t)\hat{a}_0(t)|0 \rangle $,
where $|0 \rangle$ means the initial ground state of the system, 
{\it i.e}, the state in which all the harmonic oscillators are in their 
ground states. Then we can interpret the divergent term as the particles 
of frequency $\tilde{\omega}$ produced by the modification of the initial 
ground 
state. Clearly this term does not have a thermal origin. It is a response of 
the ground state to the interaction. In computing the final
equilibrium temperature we have to use
only the part of $\bar{n}_0(t\to\infty)$ that have a thermal origin and this
corresponds to the first (finite) term in Eq. (\ref{i28}).
Alternatively note also that
the last term in Eq. (\ref{i28}) can be eliminated by making a normal
ordering of the initial annihilation and creation operators. 

The distribution given by Eq. (\ref{i28}) is one of equilibrium, but how is it
associated to a distribution of thermal equilibrium as obtained from a
thermal expectation value for the system's number operator ? 
{}For this, 
let us start by computing the thermal expectation value of the number
operator corresponding to the particle oscillator by assuming that it
is in thermal equilibrium, at some temperature $1/\beta'$, with the thermal 
bath. We then have that at thermal equilibrium the density operator for 
the total system is

\begin{equation}
\hat{\rho}_{\beta'}=\frac{e^{-\beta'\hat{H}}}{{\rm Tr}
\left(e^{-\beta'\hat{H}}\right)}\;,
\label{ap1}
\end{equation}
where $\hat{H}$ is given by Eq. (\ref{i1}) evaluated at $t\to\infty$,
that we here express it directly in its diagonal form:

\begin{eqnarray}
\hat{H}&=&\sum_{\mu=0}^N\left[\hat{A}^\dag_\mu\hat{A}_\mu+
\frac{1}{2}\right]\Omega_\mu\;,
\label{ap2}
\end{eqnarray}
where $\hat{A}^\dag_\mu$ and $\hat{A}_\mu$ are the creation and annihilation
operators associated with the normal coordinates that diagonalizes the 
original Hamiltonian and whose diagonalization transformation
between the coordinates $q_\mu$ and normal coordinates $Q_\mu$ is given by
the same matrix transformation $\eta_{\mu\nu}$, of 
Eqs. (\ref{i19}) and (\ref{i20}), 

\begin{equation}
q_\mu=\sum_{\nu=0}^N\eta_{\mu\nu}Q_\nu\;.
\label{ap3}
\end{equation}

Using Eq. (\ref{ap3}) it is easy to show that the annihilation and creation
operators, $\hat{a}_\mu$ and $\hat{a}_\mu^\dag$, associates to the coordinates
$q_\mu$, in terms of $\hat{A}^\dag_\mu$ and $\hat{A}_\mu$ are given by

\begin{equation}
\hat{a}_\mu=\sum_{\nu=0}^N\frac{\eta_{\mu\nu}}{2}
\left[\sqrt{\frac{\tilde{\omega}_\mu}{\Omega_\nu}}\left(\hat{A}_\nu+
\hat{A}_\nu^\dag\right)+\sqrt{\frac{\Omega_\nu}{\tilde{\omega}_\mu}}
\left(\hat{A}_\nu-\hat{A}_\nu^\dag\right)\right]
\label{ap4}
\end{equation}
and

\begin{equation}
\hat{a}_\mu^\dag=\sum_{\nu=0}^N\frac{\eta_{\mu\nu}}{2}
\left[\sqrt{\frac{\tilde{\omega}_\mu}{\Omega_\nu}}\left(\hat{A}_\nu+
\hat{A}_\nu^\dag\right)+\sqrt{\frac{\Omega_\nu}{\tilde{\omega}_\mu}}
\left(\hat{A}_\nu^\dag-\hat{A}_\nu\right)\right],
\label{ap5}
\end{equation}
where $\bar{\omega}_\mu=(\tilde{\omega},\omega_k)$. 
We can now compute the interacting system's equilibrium number distribution,
that we denote as $\bar{n}_0'$, which is given by

\begin{equation}
\bar{n}_0'=\frac{ {\rm Tr}\left(
\hat{a}^\dag_0\hat{a}_0~\!e^{-\beta'\hat{H}}\right)}
{{\rm Tr}(e^{-\beta'\hat{H}})}.
\label{ap6}
\end{equation}
Substituting Eq. (\ref{ap2}) and Eqs. (\ref{ap4})-(\ref{ap5}) with $\mu=0$ 
in Eq. (\ref{ap6}) we get

\begin{equation}
\bar{n}_0'=
\frac{1}{2}\sum_{\mu=0}^N\frac{\eta_{0\mu}^2}{\tilde{\omega}\Omega_\mu}
\frac{\left(\Omega_\mu^2+\tilde{\omega}^2\right)}
{\left(e^{\beta'\Omega_\mu}-1\right)}
+\frac{1}{4}\sum_{\mu=0}^N
\frac{\eta_{0\mu}^2}{\tilde{\omega}\Omega_\mu}
\left(\Omega_\mu-\tilde{\omega}\right)^2.
\label{ap7}
\end{equation}
{}Finally, by taking $N\to\infty$, $L\to\infty$ and using Eq. (\ref{i19}) in Eq.
(\ref{ap7}) we obtain, in the limit of the continuum, that

\begin{eqnarray}
\bar{n}_0'&=&\frac{2g}{\tilde{\omega}}
\int_0^\infty d\Omega\frac{\Omega(\Omega^2+\tilde{\omega}^2)}
{\left[4(\Omega^2-\tilde{\omega}^2)^2+\pi^2g^2\Omega^2\right]}
\frac{1}{\left(e^{\beta'\Omega}-1\right)}\nonumber\\
& &~~~~~~~~~~~~~~~~~~+\frac{g}{\tilde{\omega}}\int_0^\infty d \Omega
\frac{\Omega(\Omega-\tilde{\omega})^2}{4(\Omega^2-\tilde{\omega}^2)^2
+\pi^2g^2\Omega^2}.
\label{ap8}
\end{eqnarray}
Note that the above equation is identical to Eq. (\ref{i28}) if we identify
$\beta'=\beta$. Then, we have showed that the particle oscillator reachs
a final thermal equilibrium state at a temperature $T_f$ that is exactly
equal to that of the thermal bath, $T$. By
introducing dimensionless parameters, $\bar{g}=\frac{g}{\tilde{\omega}}$
and $\bar{\beta}=\tilde{\omega}\beta$, we can write for the finite part
of Eq. (\ref{i28}), that we denote as $\bar{n}_{\rm eq}$, the following 
expression

\begin{equation}
\bar{n}_{\rm eq}=2\tilde{g}
\int_0^\infty dx\frac{x(x^2+1)}
{\left[4(x^2-1)^2+\pi^2\tilde{g}^2x^2\right]
(e^{\tilde{\beta}x}-1)}\;.
\label{i29}
\end{equation}
In {}Fig. 3 we show the numerical results for $\bar{n}_{\rm eq}$, normalized
by a Bose-Einstein distribution at temperature $T$ and energy $\tilde{\omega}$,  
as a function
of scaled temperature and coupling constants, $T/\tilde{\omega}$
and $g/\tilde{\omega}$, respectively.
The results shown in {}Fig. 3 show that Eq. (\ref{i29}) quickly approaches
to the distribution $\tilde{n}_0=1/[\exp(\beta \tilde{\omega})-1]$ at high 
temperatures  $T/\tilde{\omega} >1$.    

\section{Nonlinearities at Gaussian level}

In this section we shall consider the effect of the nonlinearities at the
Gaussian approximation. The model that we consider is described by the
following Hamiltonian with quartic potential,

\begin{equation}
\hat{H}=\frac{\hat{p}^2}{2}+\frac{\omega^2(t)}{2}\hat{q}^2+
\frac{\lambda(t)}{4!}\hat{q}^4.
\label{nnq1}
\end{equation}
As in Sec. II we introduce time dependent annihilation and creation
operators,

\begin{equation}
\hat{a}(t)=A(t)\hat{q}+B(t)\hat{p},
\label{nnq2}
\end{equation}
and

\begin{equation}
\hat{a}^\dag(t)=A(t)^\ast(t)\hat{q}+B^\ast(t)\hat{p},
\label{nnq3}
\end{equation}
from which using Eq. (\ref{wrons}) we can solve for $\hat{q}$ and 
$\hat{p}$ in terms of $\hat{a}(t)$ and $\hat{a}^\dag(t)$, obtaining

\begin{equation}
\hat{q}=i\left[B^\ast(t)\hat{a}(t)-B(t)\hat{a}^\dag(t)\right]
\label{ag1}
\end{equation}
and

\begin{equation}
\hat{p}=-i\left[A^\ast(t)\hat{a}(t)-
A(t)\hat{a}^\dag(t)\right].
\label{ag2}
\end{equation}
Replacing Eqs. (\ref{ag1}) and (\ref{ag2}) in Eq. (\ref{nnq1}), normal
ordering the time dependent annihilation and creation operators and 
truncating the Hamiltonian up to quadratic terms in these operators we
obtain the result

\begin{eqnarray}
\hat{H}_G&=&\left(|A(t)|^2+\omega^2(t)|B(t)|^2+
\frac{\lambda(t)}{2}|B(t)|^4\right)\hat{a}^\dag(t)\hat{a}(t)\nonumber\\
& &~~~-\left(A^2(t)+\omega^2(t)B^2(t)+
\frac{\lambda(t)}{2}B^\star(t)B^3(t)\right)
\frac{\left[\hat{a}^\dag(t)\right]^2}{2}\nonumber\\
& &~~~~~-\left([A^\star(t)]^2+\omega^2(t)[B^\star(t)]^2+
\frac{\lambda(t)}{2}B(t)[B^\star(t)]^3\right)
\frac{\left[\hat{a}(t)\right]^2}{2}.
\label{nnq4}
\end{eqnarray}
Replacing $\hat{a}(t)$ in the LvN equation and using for $\hat{H}$ its
Gaussian approximation, as given by Eq. (\ref{nnq4}), we obtain the
following equations for $A(t)$ and $B(t)$,

\begin{equation}
A(t)=-\dot{B}(t)
\label{nnq5}
\end{equation}
and

\begin{equation}
\ddot{B}(t)+\omega^2(t)B(t)+\frac{\lambda(t)}{2}|B(t)|^2B(t)=0.
\label{nnq6}
\end{equation}
Note that now $B(t)$ is described by a nonlinear equation, that is,
this shows how at the Gaussian level approximation it is possible to 
include the nonlinearities of the model. As in Sec. II, the time dependent 
thermal
average for the number occupation operator is given by Eq. (\ref{termt}),
where the Bogoliubov coefficients are given by the same relations of
the Sec. II, the only difference is that now $B(t)$ is described by
the nonlinear equation (\ref{nnq6}). 

In general Eq. (\ref{nnq6}) cannot be solved exactly. But we can
extract some qualitative conclusions in the case in which the frequency
is time independent and when $\lambda(t)$ evolves just like as in Eq. 
(\ref{sudla}). In this case we have that

\begin{equation}
B(t\leq 0)=\frac{\rm{e}^{i\omega t}}{\sqrt{2\omega}}
\label{nnq7}
\end{equation}
and for $t\geq 0$, $B(t)$ is described by

\begin{equation}
\ddot{B}(t)+\omega^2B(t)+\frac{\lambda}{2}|B(t)|^2B(t)=0.
\label{nnq8}
\end{equation}
We have to solve the above nonlinear equation with the initial
condition (at $t=0$) given from Eq. (\ref{nnq7}), 

\begin{equation}
B=\frac{1}{\sqrt{2\omega}},~~\dot{B}=i\sqrt{\frac{\omega}{2}}.
\label{nnq9}
\end{equation}
To solve Eq. (\ref{nnq8}), it is useful to express $B(t)$ in the polar form,

\begin{equation}
B(t)=\psi(t)e^{i\phi(t)},
\label{nnq10}
\end{equation}
where $\psi(t)$ and $\phi(t)$ are real functions. Substituting Eq. 
(\ref{nnq10}) in Eq. (\ref{nnq9}) and identifying real and imaginary
parts of the resultant equation we obtain the coupled set of differential
equations

\begin{equation}
\ddot{\psi}(t)+\left(\omega^2-[\dot{\phi}(t)]^2\right)\psi(t)
+\frac{\lambda}{2}\psi^3(t)=0,
\label{nnq14}
\end{equation}

\begin{equation}
2\dot{\psi}(t)\dot{\phi}(t)+\psi(t)\ddot{\phi}(t)=0.
\label{nnq11}
\end{equation}
{}From  Eq. (\ref{nnq11}) we immediatly obtain that

\begin{equation}
\dot{\phi}(t)=c\psi^{-2}(t),
\label{nnq12}
\end{equation}
where $c$ is a constant that can be fixed by using the initial conditions. 
Comparing
Eqs. (\ref{nnq7}) and Eqs. (\ref{nnq10}), we conclude that at $t=0$,
$\psi=1/\sqrt{2\omega}$, $\dot{\psi}=0$, $\phi=0$ and 
$\dot{\phi}=\omega$. Using these values
in Eq. (\ref{nnq12}) we obtain that $c=1/2$. Then integrating Eq.
(\ref{nnq12}) with the initial condition $\phi=0$ we obtain,

\begin{equation}
\phi(t)=\frac{1}{2}\int_0^t dt'\psi^{-2}(t').
\label{nnq13}
\end{equation}
To solve for $\psi(t)$ we replace Eq. (\ref{nnq12}) (with $c=1/2$)
in Eq. (\ref{nnq14}) to obtain the result

\begin{equation}
\ddot{\psi}(t)+\omega^2\psi(t)-\frac{1}{4}\psi^{-3}(t)+
\frac{\lambda}{2}\psi^3(t)=0.
\label{nnq15}
\end{equation}
Note that the above equation is singular for $\psi=0$, but as we shall 
show below, this singularity is never reached by $\psi(t)$ in the
course of its time evolution. Using Eqs. (\ref{bug1}) and (\ref{bug2})
in Eq. (\ref{termt}) and replacing $B(t)$ as given by Eq. (\ref{nnq10}),
we obtain for the time dependent thermal expectation
value of the number operator, $\bar{n}(t)$, the following expression

\begin{eqnarray}
\bar{n}(t)&=&\left[\dot{\psi}^2(t)+\omega^2\psi^2(t)
+\frac{1}{4}\psi^{-2}(t)\right]\frac{\bar{n}}{\omega}\nonumber\\
& &~~~~~~~~~~~~~~~+\frac{1}{2}\left[\dot{\psi}^2(t)+\omega^2\psi^2(t)+
\frac{1}{4}\psi^{-2}(t)\right]
-\frac{1}{2}.
\label{nnq16}
\end{eqnarray}
Then, in order to know the time evolution of $\bar{n}(t)$ we only need
to know the time evolution of $\psi(t)$. From Eq. (\ref{nnq15}) and using
the initial conditions for $\psi(t)$ and $\dot{\psi}(t)$ we obtain

\begin{equation}
\frac{1}{2}\dot{\psi}^2(t)+\frac{\omega^2}{2}\psi^2(t)+
\frac{1}{8}\psi^{-2}(t)
+\frac{\lambda}{8}\psi^4(t)=\frac{\omega}{2}+
\frac{\lambda}{32\omega^2}.
\label{nnq17}
\end{equation}
from which we get

\begin{equation}
t=\pm\int_{1/2\omega}^{\psi^2(t)}\frac{dx}
{\sqrt{-\frac{1}{4}+\left(\omega+\frac{\lambda}{16\omega^2}\right)x
-\omega^2x^2-\frac{\lambda}{4} x^3}},
\label{nnq18}
\end{equation}
The above integral cannot be solved analytically, and then we are
not able to solve for $\psi(t)$. But we can obtain some qualitative 
conclusions about the behavior in time of $\psi(t)$ that will help
us to understand the effect of nonlinearities on the thermal
equilibration of the system we are studying.
{}For this end we rewrite
Eq. (\ref{nnq16}) by using Eq. (\ref{nnq17}), obtaining

\begin{equation}
\bar{n}(t)=\left(1+\frac{\lambda}{16\omega^3}-\frac{\lambda}{4\omega}
\psi^4(t)\right)\bar{n}+
-\frac{\lambda}{8\omega}\psi^4(t)
+\frac{\lambda}{32\omega^3}.
\label{nnq19}
\end{equation}
Then, if we know the qualitative behavior of $\psi(t)$ we would 
equally well know the qualitative behavior of $\bar{n}(t)$. If in Eq. 
(\ref{nnq17}) we think in $\psi(t)$ as being the analogous of the coordinate 
of a particle, then Eq. (\ref{nnq17}) is nothing but the energy conservation 
law of a particle of unity mass that moves in a conservative potential
$U(\psi)$ given by

\begin{equation} 
U(\psi)=\frac{\omega^2}{2}\psi^2+\frac{1}{8}\psi^{-2}+\frac{\lambda}{8}
\psi^4.
\label{nnq20}
\end{equation}
It is clear from the potential $U(\psi)$ that  
for any finite energy $\psi(t)$ evolve in time in an oscillatory way, never
reaching a fixed value, and therefore the system never reaches a final
thermal equilibrium state. Note also that in order to $\psi(t)$ to reach
the singularity $\psi=0$, the energy, in the particle mechanical 
analogy, need to be infinity. Since the energy, as given by the right
hand side of Eq. (\ref{nnq17}), is finite, we also conclude that such
singularity can  never be reached.

\section{Conclusions}

In this work we have studied several simple nonequilibrium models
represented by quantum oscillators in interaction that can be
exactly solved. In special we have studied examples of smooth
and sudden quenches that were explicitly solved in the context
of the LvN approach. As emphasized in Refs. \cite{Kim1,Kim2}, the
LvN approach has a clear advantage over other techniques to 
study out-of-thermal equilibrium problems since standard 
methods of quantum mechanics are used. 
We have explicitly seen this here by working directly in terms
of annihilation and creation operators and their Bogoliubov
transformation relations, allowing to determine the evolution
of number densities among other quantities, like the final 
equilibration temperatures and the general system time evolution.

Previous use of the method, like to the study of second
order phase transitions 
\cite{Kim2}, for the decoherence problem \cite{kim3} and for the studies
performed here, show that the LvN approach can easily be adapted
to study more complex problems that are difficult to deal with in
the context of nonequilibrium dynamics of fields where usually
the expressions for nonequilibrium propagators and correlation 
functions must be worked out. In this case the LvN approach may
provide an avenue to more rigorously take into account in the
dynamics of systems the interactions with the environment degrees of
freedom. An example of this has been worked out in details in
Sec. IV, for the model of the system quantum oscillator in interaction
with an infinity set of other quantum oscillators.

The extension of the method for applications to the current important 
problems of decoherence and description of dissipative dynamics of 
quantum systems, for the more realistic cases where  nonlinear 
interactions are present, is currently in progress and we expect to 
report on them in the near future.

\vspace{0.5cm}
\begin{center}
{\large \bf Acknowledgments}
\end{center}

R.O.R. was partially supported by Conselho Nacional de
Desenvolvimento Cient\'{\i}fico e Tecnol\'ogico - CNPq (Brazil). 
G.F.H was supported by a grant from CNPq.

\appendix

\section{}

In this appendix we solve the system of equations given by

\begin{equation}
\ddot{B}_{\mu 0}(t)+\omega_0^2B_{\mu 0}(t)+
\sum_{k=1}^N\lambda_kB_{\mu k}(t)=0,
\label{A1}
\end{equation}
\begin{equation}
\ddot{B}_{\mu k}(t)+\omega_k^2B_{\mu k}(t)+\lambda_kB_{\mu 0}(t)=0,
\label{A2}
\end{equation}
with initial conditions at $t=0$ given by

\begin{equation}
B_{\mu\nu}=\frac{\delta_{\mu\nu}}{2\omega_\mu},~~
\dot{B}_{\mu\nu}=i\delta_{\mu\nu}\sqrt{\frac{\omega_\mu}{2}}.
\label{A0}
\end{equation}
When $N=1$, $k=1$ and then we can write $\lambda_k=\lambda$ the equations 
above become identical to Eqs. 
(\ref{2eqm1})-(\ref{2eqm2}) with time independent coupling constant
(and changing the indexes $0,1$ to $1,2$). On the other 
hand, when $\lambda_k=\omega_k\sqrt{g\Delta\omega}$ we recover Eqs. 
(\ref{i8})-(\ref{i9})
with time independent coupling constant as given by Eq. (\ref{add1}). 
Also in this last case, in the initial conditions given by 
Eq. (\ref{A0}) we have to replace the bare frequencies $\omega_\mu$
by the renormalized ones 
$\tilde{\omega}_\mu=(\tilde{\omega},\omega_k)$ [see Eq. (\ref{i7})].
In order to solve Eqs. (\ref{A1})-(\ref{A2}) we introduce normal 
coordinates $Q_{\mu\nu}(t)$ through

\begin{equation}
B_{\mu\nu}(t)=\sum_{\rho=0}^N\eta_{\nu\rho}Q_{\mu\rho}(t),
\label{A3}
\end{equation}
where only the second index of $B_{\mu\nu}(t)$ appear in the matrix 
transformation $\{\eta\}$ since the first index are uncoupled in Eqs.
(\ref{A1})-(\ref{A2}). Substituting Eq. (\ref{A3}) in Eqs.
(\ref{A1})-(\ref{A2}) we obtain

\begin{equation}
\sum_{\rho=0}^N\left[\eta_{0\rho}\ddot{Q}_{\mu\rho}(t)+\omega_0^2
\eta_{0\rho}Q_{\mu\rho}(t)+\sum_{k=1}^N\lambda_k\eta_{k\rho}
Q_{\mu\rho}(t)\right]=0,
\label{A4}
\end{equation}
\begin{equation}
\sum_{\rho=0}^N\left[\eta_{k\rho}\ddot{Q}_{\mu\rho}(t)+\omega_k^2
\eta_{k\rho}Q_{\mu\rho}(t)+\lambda_k\eta_{0\rho}Q_{\mu\rho}(t)
\right]=0.
\label{A5}
\end{equation}
Multiplying Eq. (\ref{A4}) by $\eta_{0\sigma}$ and Eq. (\ref{A5}) by 
$\eta_{k\sigma}$ and adding the resulting equations we obtain

\begin{eqnarray}
&&\sum_{\rho=0}^N\left\{\left[\sum_{\nu=0}^N\omega_\nu^2\eta_{\nu\rho}
\eta_{\nu\sigma}+\sum_{k=1}^N\lambda_k\left(\eta_{k\rho}\eta_{0\sigma}
+\eta_{0\rho}\eta_{k\sigma}\right)\right]Q_{\mu\rho}(t)\right.\nonumber\\
&&~~~~~~~~~~~~~~~~~~~~~~~~~~~~~~~~~~~~~~~~~~~~~~~
+\left.\sum_{\nu=0}^N\eta_{\nu\rho}\eta_{\nu\sigma}\ddot{Q}_{\mu\rho}(t)
\right\}=0,
\label{A6}
\end{eqnarray}
and imposing that,

\begin{equation}
\sum_{\nu=0}^N\eta_{\nu\rho}\eta_{\nu\sigma}=\delta_{\rho\sigma},
\label{A7}
\end{equation}
\begin{equation}
\sum_{\nu=0}^N\omega_\nu^2\eta_{\nu\rho}\eta_{\nu\sigma}+
\sum_{k=1}^N\lambda_k\left(\eta_{k\rho}\eta_{0\sigma}+
\eta_{0\rho}\eta_{k\sigma}\right)=\delta_{\rho\sigma}\Omega_\sigma^2,
\label{A8}
\end{equation}
we obtain 

\begin{equation}
\ddot{Q}_{\mu\sigma}(t)+\Omega_\sigma^2Q_{\mu\sigma}(t)=0.
\label{A01}
\end{equation}
Eq. (\ref{A8}) can also be written as

\begin{equation}
\sum_{\nu=0}^N\left[\omega_\nu^2\eta_{\nu\rho}+\sum_{k=1}^N\lambda_k
\left(\eta_{k\rho}\delta_{0\nu}+\eta_{0\rho}\delta_{k\nu}\right)\right]
\eta_{\nu\sigma}=\delta_{\rho\sigma}\Omega_\sigma^2.
\label{A9}
\end{equation}
{}From Eqs. (\ref{A7}) and (\ref{A9}) we obtain that

\begin{equation}
\omega_\nu^2\eta_{\nu\rho}+\sum_{k=1}^N\lambda_k\left(\eta_{k\rho}
\delta_{0\nu}+\eta_{0\rho}\delta_{k\nu}\right)=\eta_{\nu\rho}
\Omega_\rho^2.
\label{A10}
\end{equation}
Taking $\nu=0$ in the above equation we obtain that 

\begin{equation}
\omega_0^2\eta_{0\rho}+\sum_{k=1}^N\lambda_k\eta_{k\rho}=\eta_{0\rho}
\Omega_\rho^2,
\label{A11}
\end{equation}
while for $\nu\neq 0$ we obtain for $\eta_{k\rho}$ the result

\begin{equation}
\eta_{k\rho}=\frac{\lambda_k}{\Omega_{\rho}^2-\omega_k^2}\eta_{0\rho}.
\label{A12}
\end{equation}
Taking $\rho=\sigma$ in Eq. (\ref{A7}) and using Eq. (\ref{A12}) 
we can solve for $\eta_{0\rho}$ obtaining the expression

\begin{equation}
\eta_{0\rho}=\left[1+\sum_{k=1}^N\frac{\lambda_k^2}{(\omega_k^2-
\Omega_\rho)^2}\right]^{-\frac{1}{2}}.
\label{A13}
\end{equation}
Using Eq. (\ref{A12}) in Eq. (\ref{A11}) we also get that

\begin{equation}
\omega_0^2-\Omega_\rho^2=\sum_{k=1}^N\frac{\lambda_k^2}{\omega_k^2
-\Omega_\rho^2}.
\label{A14}
\end{equation}
Now, from Eq. (\ref{A01}) we can write for $B_{\mu\nu}(t)$ the following
expression

\begin{equation}
B_{\mu\nu}(t)=\sum_{\rho=0}^N\eta_{\nu\rho}\left(a_{\mu\rho}
e^{i\Omega_\rho t}+b_{\mu\rho}e^{-i\Omega_\rho t}\right),
\label{A15}
\end{equation}
and using the initial conditions given by Eq. (\ref{A0}) we get for
$a_{\mu\rho}$ and $b_{\mu\rho}$, respectively, the equations

\begin{eqnarray}
&&a_{\mu\nu}=\frac{\eta_{\mu\nu}}{\sqrt{8\omega_\mu}}
\left(1+\frac{\omega_\mu}{\Omega_\nu}\right)\nonumber\\
&&b_{\mu\nu}=\frac{\eta_{\mu\nu}}{\sqrt{8\omega_\mu}}
\left(1-\frac{\omega_\mu}{\Omega_\nu}\right).
\label{A16}
\end{eqnarray}
Substituting Eq. (\ref{A16})  in Eq. (\ref{A15}) we finally obtain
the result

\begin{equation}
B_{\mu\nu}(t)=\sum_{\rho=0}^N\frac{\eta_{\mu\rho}\eta_{\nu\rho}}
{\sqrt{8\omega_\mu}}\left[\left(1+\frac{\omega_\mu}
{\Omega_\rho}\right)\rm{e}^{i\Omega_\rho t}+\left(1-\frac{\omega_\mu}
{\Omega_\rho}\right)\rm{e}^{-i\Omega_\rho t}\right].
\end{equation}

\section{}

In this section we deduce the Bogoliubov coefficients $\alpha_{00}(t)$,
$\beta_{00}(t)$, $\alpha_{0k}(t)$, and $\beta_{k0}(t)$. 
Using Eq. (\ref{i22}) in Eqs. (\ref{i23})-(\ref{i24}) we obtain

\begin{eqnarray}
\alpha_{\mu\nu}(t)&=&\sum_{\rho=0}^N\left\{
\sqrt{\frac{\tilde{\omega}_\nu}{\tilde{\omega}_\mu}}\frac{\eta_{
\mu\rho}\eta_{\nu\rho}}{4\Omega_\rho}\left[\left(\Omega_\rho+
\tilde{\omega}_\mu\right)\rm{e}^{i\Omega_\rho t}+\left(\Omega_\rho-
\tilde{\omega}_\mu\right)\rm{e}^{-i\Omega_\rho t}\right]\right.\nonumber\\
& &~~~~~~~~~~
\left.
+\frac{\eta_{\mu\rho}\eta_{\nu\rho}}{4\sqrt{\tilde{\omega}_\mu
\tilde{\omega}_\nu}}\left[\left(\tilde{\omega}_\mu+\Omega_\rho
\right)\rm{e}^{i\Omega_\rho t}+\left(\tilde{\omega}_\mu-\Omega_\rho\right)
\rm{e}^{-i\Omega_\rho t}\right]\right\},
\label{B1}
\end{eqnarray}

\begin{eqnarray}
\beta_{\mu\nu}(t)&=&\sum_{\rho=0}^N\left\{
\sqrt{\frac{\tilde{\omega}_\nu}{\tilde{\omega}_\mu}}
\frac{\eta_{\mu\rho}\eta_{\nu\rho}}{4\Omega_\rho}
\left[\left(\Omega_\rho+\tilde{\omega}_\mu
\right)\rm{e}^{i\Omega_\rho t}+\left(\Omega_\rho-\tilde{\omega}_\mu\right)
\rm{e}^{-i\Omega_\rho t}\right]\right.\nonumber\\
& &~~~~~~~~~~
\left.
-\frac{\eta_{\mu\rho}\eta_{\nu\rho}}{4\sqrt{
\tilde{\omega}_\mu\tilde{\omega}_\nu}}\left[\left(\tilde{\omega}_\mu+
\Omega_\rho
\right)\rm{e}^{i\Omega_\rho t}+\left(\tilde{\omega}_\mu-\Omega_\rho\right)
\rm{e}^{-i\Omega_\rho t}\right]
\right\}.
\label{B2}
\end{eqnarray}
Taking $\mu,\nu=0$ in the above equations we obtain

\begin{equation}
\alpha_{00}=\sum_{\rho=0}^N\frac{\left(\eta_{0\rho}\right)^2}{4}
\left[\left(2+\frac{\tilde{\omega}}{\Omega_\rho}+\frac{\Omega_\rho}
{\tilde{\omega}}\right)e^{i\Omega_\rho t}+\left(2-\frac{\tilde{\omega}}
{\Omega_\rho}-\frac{\Omega_\rho}{\tilde{\omega}}\right)e^{-i
\Omega_\rho t}\right],
\label{B3}
\end{equation}

\begin{equation}
\beta_{00}=\sum_{\rho=0}^N\frac{\left(\eta_{0\rho}\right)^2}{4}\left[
\left(\frac{\tilde{\omega}}{\Omega_\rho}-\frac{\Omega_\rho}{\tilde{\omega}}
\right)e^{i\Omega_\rho t}+\left(\frac{\Omega_\rho}
{\tilde{\omega}}-\frac{\tilde{\omega}}{\Omega_\rho}\right)
e^{-i\Omega_\rho t}\right].
\label{B4}
\end{equation}
Substituting Eq. (\ref{i19}) in the above equations and going to the continuum 
limit we obtain, {\it e.g.} for $\alpha_{00}(t)$, the 
expression

\begin{eqnarray}
\alpha_{00}(t)&=&\frac{g}{\tilde{\omega}}
\int_{0}^{\infty} d\Omega\frac{\Omega\left[
(\tilde{\omega}+\Omega)^2e^{i\Omega t}-(\tilde{\omega}-\Omega)^2
e^{-i\Omega t}\right]}
{4\left(\Omega^2-\tilde{\omega}^2\right)^2+\pi^2g^2\Omega^2}
\nonumber\\
&=&\frac{g}{2\tilde{\omega}}
\int_{-\infty}^{\infty} d\Omega\frac{\Omega\left[
(\tilde{\omega}+\Omega)^2e^{i\Omega t}-(\tilde{\omega}-\Omega)^2
e^{-i\Omega t}\right]}
{4\left(\Omega^2-\tilde{\omega}^2\right)^2+\pi^2g^2\Omega^2}\nonumber\\
&=&\frac{g}{\tilde{\omega}}\int_{-\infty}^{\infty} 
d\Omega\frac{\Omega(\tilde{\omega}+\Omega)^2}
{4\left(\Omega^2-\tilde{\omega}^2\right)^2+\pi^2g^2\Omega^2}
e^{i\Omega t}.
\label{B5}
\end{eqnarray}
The integral in Eq. (\ref{B5}) can be easily evaluated 
using the Cauchy theorem and
choosing as contour of integration in the upper half plane. The poles
located in the upper half plane are $\pm\tilde{\tilde{\omega}}
+\frac{i\pi g}{4}$, where $\tilde{\tilde{\omega}}=\sqrt{\tilde{\omega}^2
-\frac{\pi^2g^2}{16}}$. Assuming $\tilde{\omega}>\frac{\pi g}{4}$ the 
result is

\begin{equation}
\alpha_{00}(t)=\frac{e^{-\pi g t/4}}{64\tilde{\omega}
\tilde{\tilde{\omega}}}\left[\left(4\tilde{\omega}+
4\tilde{\tilde{\omega}}+i\pi g\right)^2
e^{i\tilde{\tilde{\omega}}t}-
\left(4\tilde{\omega}-4\tilde{\tilde{\omega}}+i\pi g\right)^2
e^{-i\tilde{\tilde{\omega}}t}\right].
\label{B6}
\end{equation}
In the same way we obtain from Eq. (\ref{B4}) the expression
for $\beta_{00}(t)$
(assuming again $\tilde{\omega}>\frac{\pi g}{4}$),

\begin{equation}
\beta_{00}(t)=\frac{\pi ge^{-\pi g t/4}}{32\tilde{\omega}
\tilde{\tilde{\omega}}}\left[\left(\pi g-4i\tilde{\tilde{\omega}}
\right)e^{i\tilde{\tilde{\omega}}t}-
\left(\pi g+4i\tilde{\tilde{\omega}} \right)
e^{-i\tilde{\tilde{\omega}}t}\right].
\label{B7}
\end{equation}

Setting $\mu=0,~\!\nu=k$ in Eq. (\ref{B1}) we obtain,

\begin{eqnarray}
\alpha_{0k}(t)&=&\frac{\omega_k g\sqrt{g\Delta \omega}}
{\sqrt{\tilde{\omega}\omega_k}}
\sum_\rho \frac{\Delta \omega\Omega}{\Omega_\rho^2-\omega_k^2}
\left[\frac{(\omega_k+\Omega_\rho)(\Omega_\rho+
\tilde{\omega})e^{i\Omega_\rho t}}
{4\left(\Omega^2-\tilde{\omega}^2\right)^2+\pi^2g^2\Omega^2}\right.
\nonumber\\
& &~~~~~~~~~~~~~~~~~~~~~~~~~~~~~~~~~~~
\left.+\frac{(\omega_k-\Omega_\rho)(\Omega_\rho-\tilde{\omega})
e^{-i\Omega_\rho t}}
{4\left(\Omega^2-\tilde{\omega}^2\right)^2+\pi^2g^2\Omega^2}
\right].
\label{B8}
\end{eqnarray}
In going to the continuum limit we have to take care since
there will appear a singularity when $\Omega=\omega_k$. In reality
this singularity is fictitious and it can be eliminated from Eq.
(\ref{B8}) using Eq. (\ref{i17}). In terms of the renormalized
frequency $\tilde{\omega}$ Eq. (\ref{i17}) can be written as
(setting $\mu=k$)

\begin{eqnarray}
\tilde{\omega}^2-\Omega_k^2&=&\sum_{q\neq k}\frac{g\Omega_k^2\Delta\omega}
{\omega_q^2-\Omega_k^2}+
\frac{g\Omega_k^2\Delta \omega}{\omega_k^2-\Omega_k^2}\nonumber\\
&=&g\int_0^\infty d'\omega\frac{\Omega_k^2}{\omega^2-\Omega_k^2}
+\frac{g\Omega_k^2\Delta \omega}{\omega_k^2-\Omega_k^2},
\label{B9}
\end{eqnarray}
where we have isolated the pole at $\omega_k=\Omega_k$ and the prime
in the integral means that the point $\omega=\Omega_k$ has been
excluded from the integration. It is easy to show that such
integration is zero. For this end we write such integration as

\begin{equation}
I=\frac{g}{2}\int_{-\infty}^{\infty}d'\omega\frac{\Omega_k^2}
{\omega^2-\Omega_k^2},
\label{B10}
\end{equation}
where now it is understood that we omit the points $\pm\Omega_k=\omega_k$
in doing the integration over the frequency. 
Going to the complex plane and using Cauchy
theorem we obtain,

\begin{equation}
I=\frac{g\Omega_k^2}{2}\left(\frac{i\pi}{2\Omega_k}+
\frac{i\pi}{-2\Omega_k}\right)=0.
\label{11}
\end{equation}
Then, from Eq. (\ref{B9}) we obtain

\begin{equation}
\tilde{\omega}^2-\Omega_k^2=
\frac{g\Omega_k^2\Delta \omega}{\omega_k^2-\Omega_k^2},
\label{B12}
\end{equation}
where it is understood that $\Omega_k=\omega_k$ and since we are 
considering the continuum
limit, in which case $\Delta \omega \to 0$, then the 
right-hand-side of Eq. (\ref{B12}) is well defined. Now we can use
Eq. (\ref{B12}) in Eq. (\ref{B8}) to eliminate the singularity after
isolating it. Isolating the singularity in Eq. (\ref{B8}) and going to
the continuum limit we obtain

\begin{eqnarray}
\alpha_{0k}(t)&=&\frac{\omega_k g\sqrt{g\Delta \omega}}
{\sqrt{\tilde{\omega}\omega_k}}
\int_0^\infty d' \Omega\frac{\Omega}{\Omega_\rho^2-\omega_k^2}
\left[\frac{(\omega_k+\Omega)(\Omega+\tilde{\omega})
e^{i\Omega t}}{4\left(\Omega^2-\tilde{\omega}^2\right)^2
+\pi^2g^2\Omega^2}\right.\nonumber\\
& &
\left.~~~+\frac{(\omega_k-\Omega)(\Omega-\tilde{\omega})
e^{-i\Omega t}}{4\left(\Omega^2-\tilde{\omega}^2\right)^2
+\pi^2g^2\Omega^2}
\right]
+\frac{\omega_k g\sqrt{g\Delta \omega}}
{\sqrt{\tilde{\omega}\omega_k}}
\frac{\Delta \omega\Omega_k}{\Omega_k^2-\omega_k^2}
\nonumber\\
& &~~~\times\left[
\frac{(\omega_k+\Omega_k)(\Omega_k+
\tilde{\omega})e^{i\Omega_k t}}
{4\left(\Omega_k^2-\tilde{\omega}^2\right)^2+\pi^2g^2\Omega_k^2}+
\frac{(\omega_k-\Omega_k)(\Omega_k-\tilde{\omega})
e^{-i\Omega_kt}}
{4\left(\Omega_k^2-\tilde{\omega}^2\right)^2+\pi^2g^2\Omega_k^2}
\right]\nonumber\\
&=&g\sqrt{\frac{\omega_k}{\tilde{\omega}}}\sqrt{g\Delta\omega}
\int_{-\infty}^\infty d'\Omega\frac{\Omega(\Omega+\tilde{\omega})
e^{i\Omega t}}{(\Omega-\omega_k)
\left[4\left(\Omega^2-\tilde{\omega}^2\right)^2+\pi^2g^2\Omega^2\right]}
\nonumber\\
& &~~~~~~~~~~~~~~~~~~~~~~~
-2\frac{\sqrt{g\Delta \omega}\omega_k(\omega_k+
\tilde{\omega})(\tilde{\omega}^2-\omega_k^2)e^{i\omega_k t}}
{\sqrt{\tilde{\omega}\omega_k}\left[4(\omega_k^2-\tilde{\omega}^2)^2
+\pi^2g^2\omega_k^2\right]},
\label{B13}
\end{eqnarray}
where in passing to the second equality we have used the even symmetry of 
the integrand, factored the term $(\omega_k+\Omega)$, also we have used
Eq. (\ref{B12}) to eliminate the singularity 
$(\Omega_k^2-\omega_k^2)^{-1}$ and finally we set $\Omega_k=\omega_k$.
The prime in the integrand of Eq. (\ref{B13}) means that we are
avoiding the points $\pm\omega_k$. Then when using the Cauchy theorem we
have to multiply the residues in these poles by $i\pi$ instead of $2i\pi$.
Actually the pole at $-\omega_k$ has been factored. Then after using Cauchy
theorem it is easy to obtain,

\begin{eqnarray}
\alpha_{0k}(t)&=&\sqrt{\frac{\omega_k}{\tilde{\omega}}}
\frac{(\tilde{\omega}+\omega_k)(2\omega_k^2-2\tilde{\omega}^2+i\pi g
\omega_k)}{\left[4\left(\omega_k^2-\tilde{\omega}^2\right)^2+
\pi^2g^2\omega_k^2\right]}
\sqrt{g\Delta \omega}\,e^{i\omega_k t}
+\sqrt{\frac{\omega_k}{\tilde{\omega}}}\frac{\sqrt{g\Delta\omega}}
{4\tilde{\tilde{\omega}}}\nonumber\\
& &
\times \!\!\left[\frac{(4\tilde{\tilde{\omega}}+4\tilde{\omega}+i\pi g)}
{(4\tilde{\tilde{\omega}}-4\omega_k+i\pi g)}
e^{i\tilde{\tilde{\omega}}t}
+\frac{(4\tilde{\omega}-4\tilde{\tilde{\omega}}+i\pi g)}
{(4\tilde{\tilde{\omega}}+4\omega_k-i\pi g)}
e^{-i\tilde{\tilde{\omega}}t}\right]e^{-\pi g t/4}.
\label{B14}
\end{eqnarray}

Setting $\mu=k,~\nu=0$ in Eq. (\ref{B2}) we obtain an expression
for $\beta_{k0}(t)$ that contains also the same singularity as in the
case of $\alpha_{0k}(t)$. Then, proceeding in the same way as in the
above case we obtain

\begin{eqnarray}
\beta_{k0}(t)&=&\sqrt{\frac{\omega_k}{\tilde{\omega}}}
\frac{(\tilde{\omega}-\omega_k)(2\omega_k^2-2\tilde{\omega}^2+i\pi g
\omega_k)}{\left[4\left(\omega_k^2-\tilde{\omega}^2\right)^2+
\pi^2g^2\omega_k^2\right]}
\sqrt{g\Delta \omega}\,e^{i\omega_k t}
+\sqrt{\frac{\omega_k}{\tilde{\omega}}}
\frac{\sqrt{g\Delta\omega}}{4\tilde{\tilde{\omega}}}\nonumber\\
& &
\times\!\!
\left[\frac{(4\tilde{\omega}-4\tilde{\tilde{\omega}}-i\pi g)}
{(4\tilde{\tilde{\omega}}-4\omega_k+i\pi g)}
e^{i\tilde{\tilde{\omega}}t}+
\frac{(4\tilde{\omega}+4\tilde{\tilde{\omega}}-i\pi g)}
{(4\tilde{\tilde{\omega}}+4\omega_k-i\pi g)}
e^{-i\tilde{\tilde{\omega}}t}\right]e^{-\pi g t/4}.
\label{B15}
\end{eqnarray}

\newpage

\begin{figure}[c] 
\epsfysize=10cm  
{\centerline{\epsfbox{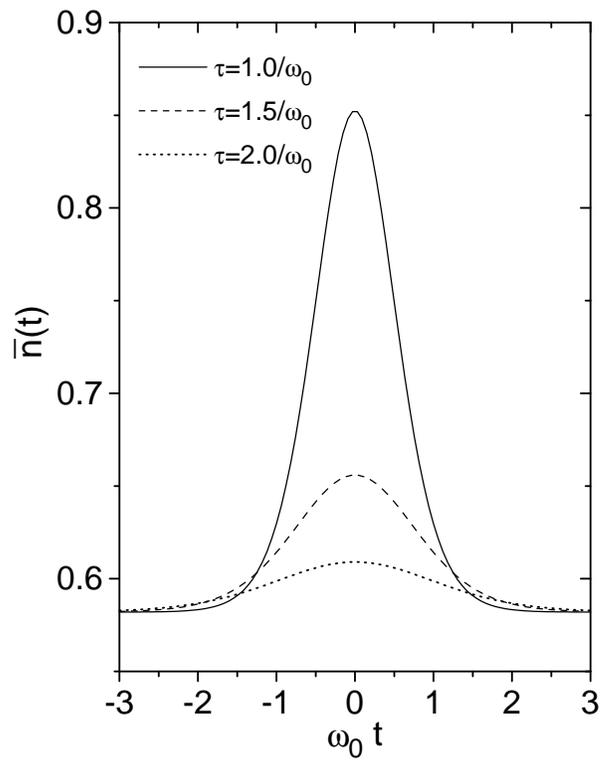}}}
\caption{The number density for $\omega_0 \beta=1$ and for
decay times (in units of $1/\omega_0$) $\tau =1.0,\;1.5$ and $2.0$.}

\end{figure}

\begin{figure}[c] 
\epsfysize=10cm  
{\centerline{\epsfbox{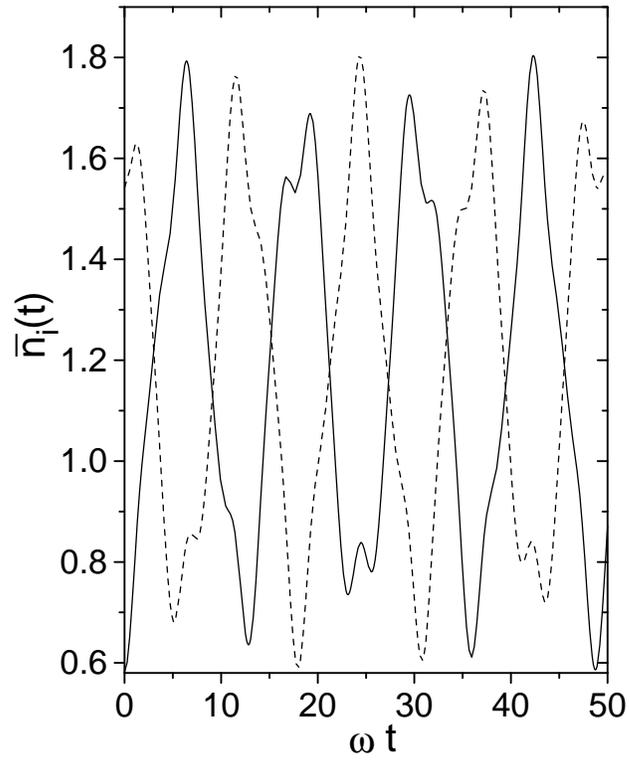}}}
\caption{The time evolution for the number densities $\bar{n}_1 (t)$
(dashed curve) and $\bar{n}_2 (t)$ (continuous curve) for $\omega_1=
\omega_2=\omega$, $\omega\beta_1=1/2$, $\omega\beta_2=1$ and $\lambda
=\omega^2/2$.}

\end{figure}

\begin{figure}[c] 
\epsfysize=10cm  
{\centerline{\epsfbox{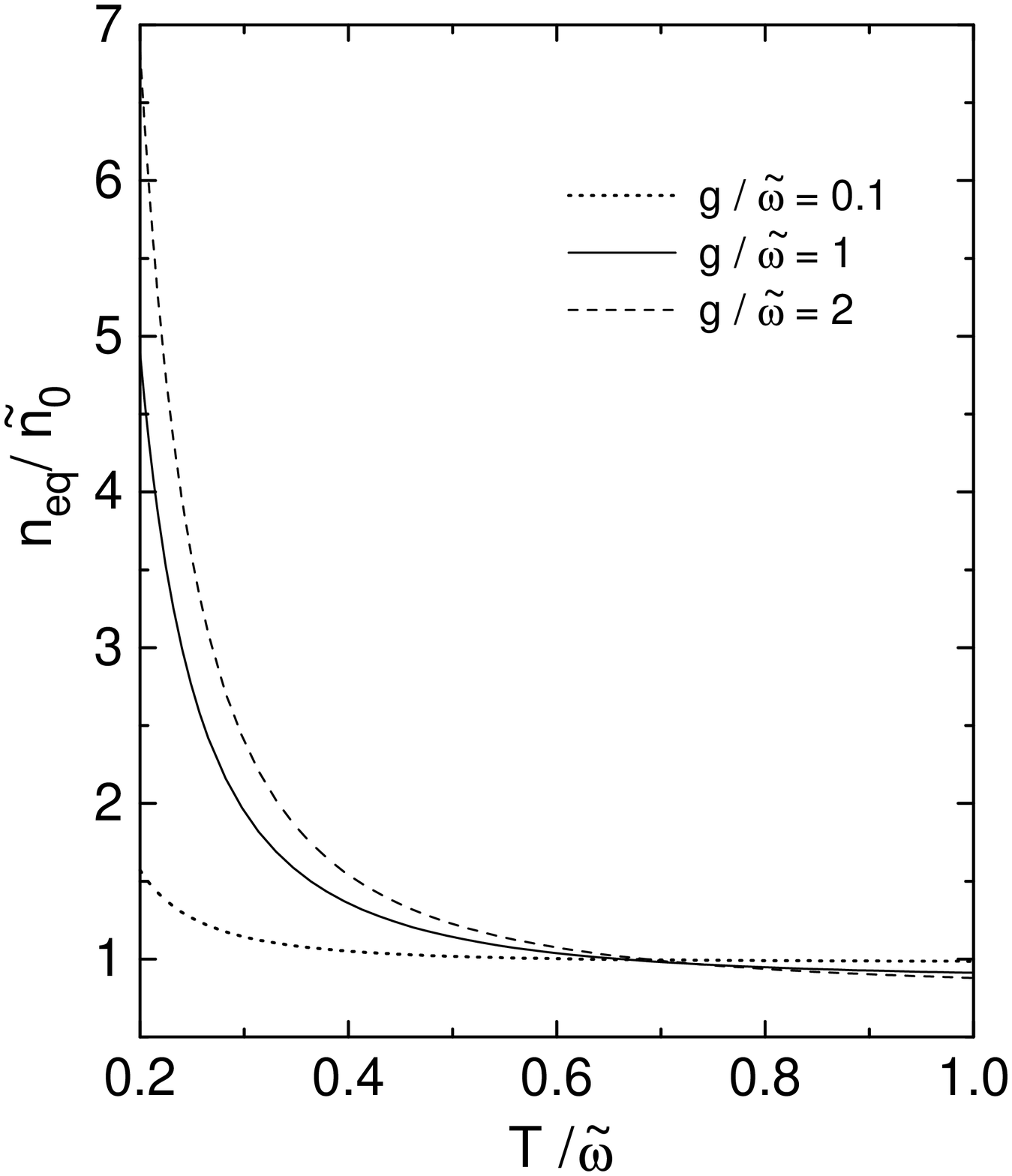}}}
\caption{The system's equilibrium distribution, Eq. (\ref{i29}), 
normalized by the distribution $\tilde{n}_0=1/[\exp(\beta\tilde{\omega})-1]$,
as a function of the (scaled) temperature $T$ and
for different values of the coupling constant $g$.}

\end{figure}


\begin{thebibliography}{99}

\bibitem{reviews} J. S. Langer in {\it Solids Far from Equilibrium}, Ed. C.
Godr\`eche (Cambridge University Press, Cambridge, 1992); D. J. Evans and G. P.
Morriss, {\it Statistical Mechanics of Non-Equilibrium Liquids} 
(Academic Press, London, 1990).

\bibitem{DCC} K. Rajagopal and F. Wilczek, Nucl. Phys. {\bf B 399}, 395 (1995);
ibid. {\bf B 404} (1993) 577.

\bibitem{lindereh}L. Kofman, A. Linde and A. A. Starobinskii, 
Phys. Rev. {\bf D 56} (1997) 3258.

\bibitem{GR}M. Gleiser and R. O. Ramos, Phys. Rev. {\bf D 50} (1994) 2441.

\bibitem{BGR}A. Berera, M. Gleiser and R. O. Ramos, Phys. Rev. {\bf D 58}
(1998) 123508.

\bibitem{BR}A. Berera and R. O. Ramos, Phys.
Rev. {\bf D 63} (2001) 103509.

\bibitem{BEC}D. G. Barci, E. S. Fraga and R. O. Ramos,
Phys. Rev. Lett. {\bf 85} (2000) 479; 
Laser Phys.  {\bf 12} (2002) 43;
D. G. Barci, E. S. Fraga, M. Gleiser and R. O. Ramos,
Physica {\bf A 317} (2003) 535.

\bibitem{RF}R. O. Ramos and F. A. R. Navarro, Phys. Rev. {\bf D 62}
(2000) 085016.

\bibitem{ctpreviews}K. Chou, Z. Su, B. Hao and L. Yu, Phys. Rep.
{\bf 118} (1985) 1;
N. P. Landsman and Ch. G. van Weert, Phys. Rep.
{\bf 145} (1987) 141.

\bibitem{Lewis}H. R. Lewis, Jr. and W. B. Riesenfeld, J. Math.
Phys. {\bf 10} (1969) 1458.

\bibitem{Kim1}S. P. Kim and C. H. Lee, Phys. Rev. {\bf D 62}
(2000) 125020. 

\bibitem{Kim2}S. P. Kim, S. Sengupta and F. C. Khanna,
Phys. Rev. {\bf D 64} (2001) 105026. 

\bibitem{solitons} P. G. Drazin and R. S. Johnson, {\it Solitons:
An Introduction} (Cambridge University Press, Cambridge, England, 1993).

\bibitem{gradshteyn} I. S. Gradshteyn and L. M. Ryzhic, {\it Table
of Integrals Series and Products} (Academic Press, Inc., New York,
1980).

\bibitem{jackiw}O. \'Eboli, R. Jackiw and S.-Y. Pi, Phys.
Rev. {\bf D 37} (1988) 3557.

\bibitem{abramov} M. Abramowitz and I. Stegun, {\it Handbook of Mathematical
Functions} (Dover Publications, Inc., New York, 1964).

\bibitem{caldeira}A. O. Caldeira and A. J. Leggett, Ann. Phys. (N.Y.)
{\bf 149} (1983) 374.

\bibitem{adolfo}N. Andion, A. P. C. Malbouisson and A. Mattos Neto, 
J. Phys. {\bf A 34} (2001) 3735.

\bibitem{kim3}S. P. Kim and C. H. Lee, Phys. Rev. {\bf D 65} 
(2002) 045013.


\end{thebibliography}
\end{document}